\documentclass[aps,prb,twocolumn,groupedaddress,floatfix]{revtex4-1}
\usepackage{hyperref}
\usepackage{epsfig}
\usepackage{amssymb,amsmath}
\usepackage{tikz}
\usetikzlibrary{decorations.pathreplacing}
\usepackage{standalone}
\usetikzlibrary{calc}
\usepackage{verbatim}
\usepackage{float}
\usepackage{lipsum,booktabs}
\usepackage{wrapfig,yfonts}
\usepackage{titlesec}
\usepackage{dcolumn,marvosym}
\usepackage{enumitem}

\usepackage{mathtools}
\DeclarePairedDelimiter\bra{\langle}{\rvert}
\DeclarePairedDelimiter\ket{\lvert}{\rangle}
\DeclarePairedDelimiterX\braket[2]{\langle}{\rangle}{#1 \delimsize\vert #2}
\newcommand*\circled[1]{\tikz[baseline=(char.base)]{
\node[shape=circle,draw,inner sep=0.5pt] (char) {#1};}}

\def\swne{
    \tikz[baseline=0.4ex]{
        \draw (0,0) -- (0,2ex) -- (2ex,2ex) -- (2ex,0) -- (0,0);
        \fill (0,0) circle (2pt);
        \fill (2ex,2ex) circle (2pt);
        \draw[fill=white] (0,2ex) circle (2pt);
        \draw[fill=white] (2ex,0) circle (2pt);
    }
}
\def\nwse{
    \tikz[baseline=0.4ex]{
        \draw (0,0) -- (0,2ex) -- (2ex,2ex) -- (2ex,0) -- (0,0);
        \fill (0,2ex) circle (2pt);
        \fill (2ex,0) circle (2pt);
        \draw[fill=white] (2ex,2ex) circle (2pt);
        \draw[fill=white] (0,0) circle (2pt);
    }
}

\begin{document}
\title{Entanglement Signatures of Multipolar Higher Order Topological Phases}
\date{\today}

\author{Oleg Dubinkin}
\email[]{olegd2@illinois.edu}
\author{Taylor L. Hughes}
\email[]{hughest@illinois.edu}

\affiliation{Department of Physics and Institute for Condensed Matter Theory, University of Illinois at Urbana-Champaign, Urbana, IL, 61801-3080, USA}

\begin{abstract}
  We propose a procedure that characterizes free-fermion or interacting multipolar higher-order topological phases via their bulk entanglement structure. To this end, we construct \emph{nested} entanglement Hamiltonians by first applying an entanglement cut to the ordinary many-body ground state, and then iterating the procedure by applying further entanglement cuts to the (assumed unique) ground state of the \emph{entanglement} Hamiltonian. We argue that an $n$-th order multipolar topological phase can be characterized by the features of its $n$-th order \textit{nested entanglement Hamiltonian} e.g., degeneracy in the entanglement spectrum. We explicitly compute nested entanglement spectra for a set of higher-order fermionic and bosonic multipole phases and show that our method successfully identifies such phases. 
\end{abstract}

\pacs{spin, SPT, Heisenberg, Quadrupole}

\maketitle

\section{Introduction}
The presence of protected gapless modes on the $(d-1)$-dimensional edge of a gapped $d$-dimensional system is one of the key signatures of topologically non-trivial phases of matter. 
Recently this paradigm was expanded to include higher-order symmetry protected topological (HOSPT) phases in fermionic\cite{quadrupole,quadrupole2,hoti2018,Trifunovic2018,Song2017,Wang2018,Langbehn2017,teo2013,Isobe15,benalcazar2014,Song16,benalcazar2018,varjas19,dubinkin2019dipole} and bosonic\cite{You18,Dubinkin18,YouDevakul18,youburnellhughes2019} systems.
A feature that unifies all of these phases is the existence of symmetry-protected topological features, such as zero modes or fractional charge, localized at the subdimensional boundaries of the lattice (e.g., at corners or hinges where multiple surfaces intersect). 

Upon closer inspection, a particular subset of higher order models shares a rich hierarchical spatial structure. This structure was first demonstrated in the quadrupole topological insulator\cite{quadrupole} (QTI) where the topological quadrupole phase, having non-vanishing bulk quadrupole moment $q_{xy},$ is protected by a pair of anticommuting mirror symmetries, or by $C_4$ symmetry. The consequences of a bulk quadrupole moment can be seen by taking a square shaped sample having open boundaries. In such a sample, two edges, parallel to $\hat{x},$ and $\hat{y}$, respectively, that intersect at a corner can be polarized, and the system generically obeys the equation
\begin{equation}
    Q^{corner}-P_{1}^{edge}-P_{2}^{edge}=-q_{xy},
    \label{eqn:indicies_pol}
\end{equation} where $P^{edge}_{1,2}$ are the edge polarizations for the two intersecting edges, $Q^{corner}$ is the corner charge, and $q_{xy}$ is the bulk quadrupole moment.
Interestingly, a similar relationship was shown to hold between the $\mathbb{Z}_2$ indices of lower dimensional SPT phases appearing at the edges of a 2D bosonic, second order quadrupolar SPT phase\cite{Dubinkin18}, and the protected  degrees of freedom localized at the corner. This observation suggests that Eq. (\ref{eqn:indicies_pol}) can be understood as a type of quadrupolar bulk-boundary correspondence that relates the topological indices of a hierarchy of topological phenomena in descending dimensions, and defines a certain subclass of second order SPT phases. 

An important aspect of this concept is that the edges of HOSPTs are gapped themselves, and in the presence of residual symmetry, can harbor a 1D first order SPT phase. This idea was further investigated in Refs. \onlinecite{quadrupole2,khalaf19} where the notion of boundary obstructions was introduced for free-fermion models and characterized using the spectrum of the \emph{Wannier Hamiltonian}, which shares some of its spectral structure with the Hamiltonian at a physical edge\cite{fidkowski11,quadrupole}.
Moreover, computing topological invariants constructed from a subset of the gapped edge/Wannier bands yields a refined topological classification that can identify some types of higher order topology in non-interacting band theories. 

Using the topological properties of the Wannier Hamiltonian bands as a bulk proxy for the edge Hamiltonian has proved  fruitful when examining lattice models of free fermions. 
However, this approach is no longer applicable once interactions are taken into account. 
Seeing that both free and interacting models of HOSPT phases share many features and, in some cases, can be mapped onto each other\cite{Dubinkin18}, it is natural to search for a more generically applicable method that will allow the identification of quadrupolar-like HOSPTs from the bulk many-body wave function alone.

To make further progress let us turn our attention to the entanglement structure of the ground state, which is known to serve as a bulk indicator of topological edge properties\cite{LiHaldane}.
In particular, a relevant result for our work is that one obtains a protected, degenerate entanglement spectrum when cutting non-trivial 1D SPT phases\cite{ptbo}. More so, the entanglement Hamiltonian itself has been shown to be in direct correspondence with the Hamiltonian of the physical edge\cite{Qi12,sterdyniak2012,Swingle12}. Hence, for HOSPTs we would expect to find that the entanglement spectrum  would have a unique, gapped ground state since the edge spectrum is gapped.
Indeed, quite recently, the entanglement Hamiltonian computed for the ground state of the QTI model was shown\cite{fukui18} to reproduce all of the crucial features of the physical edge, allowing one to relate the gap in the entanglement spectrum and the Berry phase/polarization of the entanglement ground state with the corresponding quantities at the physical edge. The entanglement spectrum has also been used to characterize some 3D HOSPTs as well\cite{hoti2018}.

These observations inspire us to further investigate the applicability of entanglement Hamiltonians to access the topological properties of some classes of many-body HOSPT systems. 
Specifically, we define a characterization using a series of \emph{Nested Entanglement Hamiltonians} (NEH) which, as we show, can be used as a tool to access the essential boundary physics purely from the ground state wave function of a completely periodic $d$-dimensional system. 
As described in detail in Section \ref{NEH_method}, we perform a series of $n$ entanglement cuts that intersect on a $(d-n)$-dimensional subspace. We begin with the ground state of the physical Hamiltonian and make an entanglement cut. We can then study the associated entanglement Hamiltonian and its spectrum. If it is gapped, or can be gapped with suitable symmetry-preserving perturbations, we can then extract the entanglement ground state for the first entanglement Hamiltonian. Assuming this ground state does not spontaneously break the protective symmetries, we can perform a second cut etc. Hence, we define an $n$-th order NEH recursively as an entanglement Hamiltonian computed for a spatial cut of the ground state of the $(n-1)$-th entanglement Hamiltonian. 
The ground state of the physical Hamiltonian serves as a base for this recursive procedure.

In this article we illustrate this characterization procedure using a series of non-interacting and interacting fermionic and bosonic models. 
By studying a class of HOSPT models that maintain some residual symmetry at the boundary, we show that the entanglement cuts either produce a gapped entanglement Hamiltonian that hosts an SPT phase, or a gapless entanglement Hamiltonian that hosts protected degenerate states.
This observation prompts us to consider entanglement features, e.g., protected degeneracy of the $n$-th nested entanglement Hamiltonian ground state, along with the uniqueness of the ground states of every $(n-k)$-th NEH for $k=1,2...(n-1),$ as a characteristic of some HOSPT phases that can be applied to free and interacting systems. For the models we study here, the nested entanglement Hamiltonians and their low-lying spectral properties provide useful characterizations of the associated HOSPT phases. For more generic models, especially in 3D systems where other subtleties in the nested 2D spectra can arise\cite{cano2015,santos2018,williamson2019}, one may have to consider additional features of the entanglement spectrum to find a completely generic entanglement classification scheme for HOSPTs (see Ref. \onlinecite{you2020entanglement} for some further discussion).

Our article is organized as follows. We begin with a detailed presentation of our method in Section \ref{NEH_method}.
Then we proceed to Section \ref{results} where we apply our proposed construction to a variety of lattice models that are  known to develop quadrupolar and octupolar topological phases in the ground state. We explicitly compute eigenspectra of their second and third order nested entanglement Hamiltonians  and show that these spectra turn out to be degenerate for the non-trivial quadrupolar and octupolar HOSPT phases respectively. Finally, we conclude in Section \ref{sec:conclusion}.

\begin{figure}
  \includegraphics[width=0.3\textwidth]{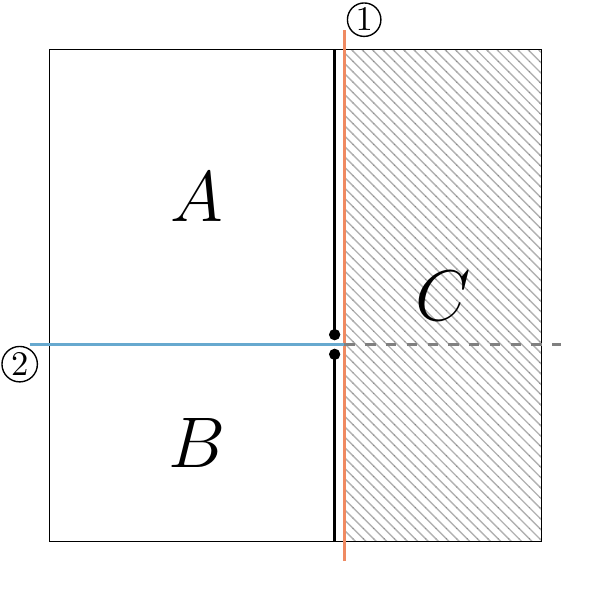}
\caption{A two-dimensional system with two successive entanglement cuts. The first cut shown in red results in a gapped SPT phase at the entanglement edge (black line), the second cut shown in blue ends up splitting the SPT phase leading to a doubly-degenerate entanglement spectrum.}
\label{fig:2d_cuts}
\end{figure}

\section{Method of Nested Entanglement Hamiltonians}
\label{NEH_method}
In this section we will describe the notion of a Nested Entanglement Hamiltonian (NEH) and what information we can expect to extract from some types of HOSPT phases. 
We can define the $(n+1)$-th NEH, denoted as $H^{E,n+1}$, recursively: $H^{E,n+1}$ is the entanglement Hamiltonian computed from a bi-partitioned reduced density matrix of the ground state (assumed to be non-degenerate) of the $n$-th NEH $H^{E,n}.$ 
The resulting reduced density matrix $\rho^{n+1}$ encodes the $(n+1)$-th NEH via $\rho^{n+1}=\exp(-H^{E,n+1})$.
The base of this recursive procedure is the physical Hamiltonian of the system, i.e.,  $H^{E,0}\equiv H$. For the purpose of characterizing $d$-dimensional HOSPTs we choose that the set of $n$ spatial cuts used to generate the series of NEHs, are represented by surfaces of co-dimension 1 that have a non-trivial intersection over a $(d-n)$ dimensional manifold. For instance, to define a 3rd order NEH in 3 spatial dimensions we need to consider three consecutive planar cuts that all have at least one common point. In practice, we will consider lattice models where  entanglement cuts are represented by a series of co-dimension 1 planes. 

To briefly motivate the usefulness of this construction, and to set our expectations for the detailed examples in the next section, let us imagine a 2-dimensional quadrupolar HOSPT with the Hamiltonian $H$. To construct the series of NEHs, we first we make an entanglement cut $\circled{1}$ (shown in Fig. \ref{fig:2d_cuts}) and compute the reduced density matrix, and corresponding entanglement Hamiltonian, for the ground state $\ket{\Psi^{(0)}}$ of $H$: 
\begin{equation}
    \rho^{(1)}_{AB}=\text{Tr}_{C}\rho^{(0)}=\exp\left(-H^{E,1}_{AB}\right),
    \label{eqn:def_ent_ham}
\end{equation}
where $\rho^{(0)}=\ket{\Psi^{(0)}}\bra{\Psi^{(0)}}$ is the full ground state density matrix, and $\rho^{(1)}_{AB}$ is the reduced density matrix for the entanglement cut that separates region $AB$ from region $C$ (see Fig. \ref{fig:2d_cuts}).

\begin{figure}
\includegraphics[width=0.25\textwidth]{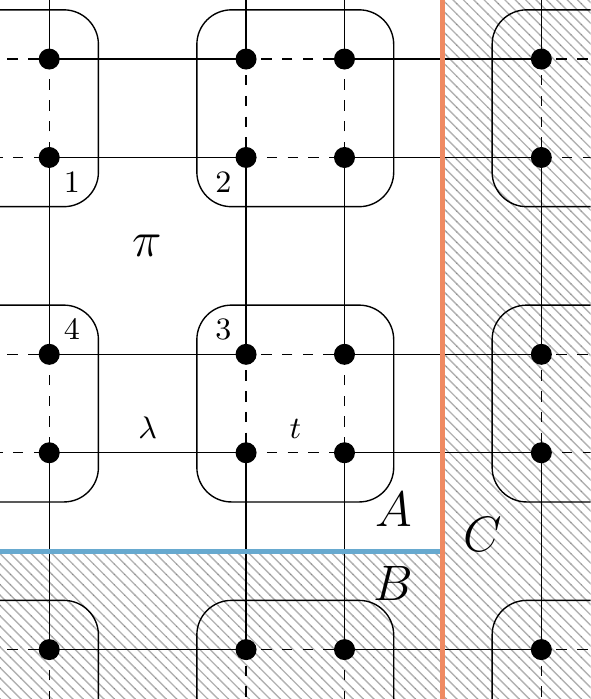}
\begin{tikzpicture}
  \node[anchor=south west,inner sep=0] at (0,0) {\includegraphics[width=0.2\textwidth]{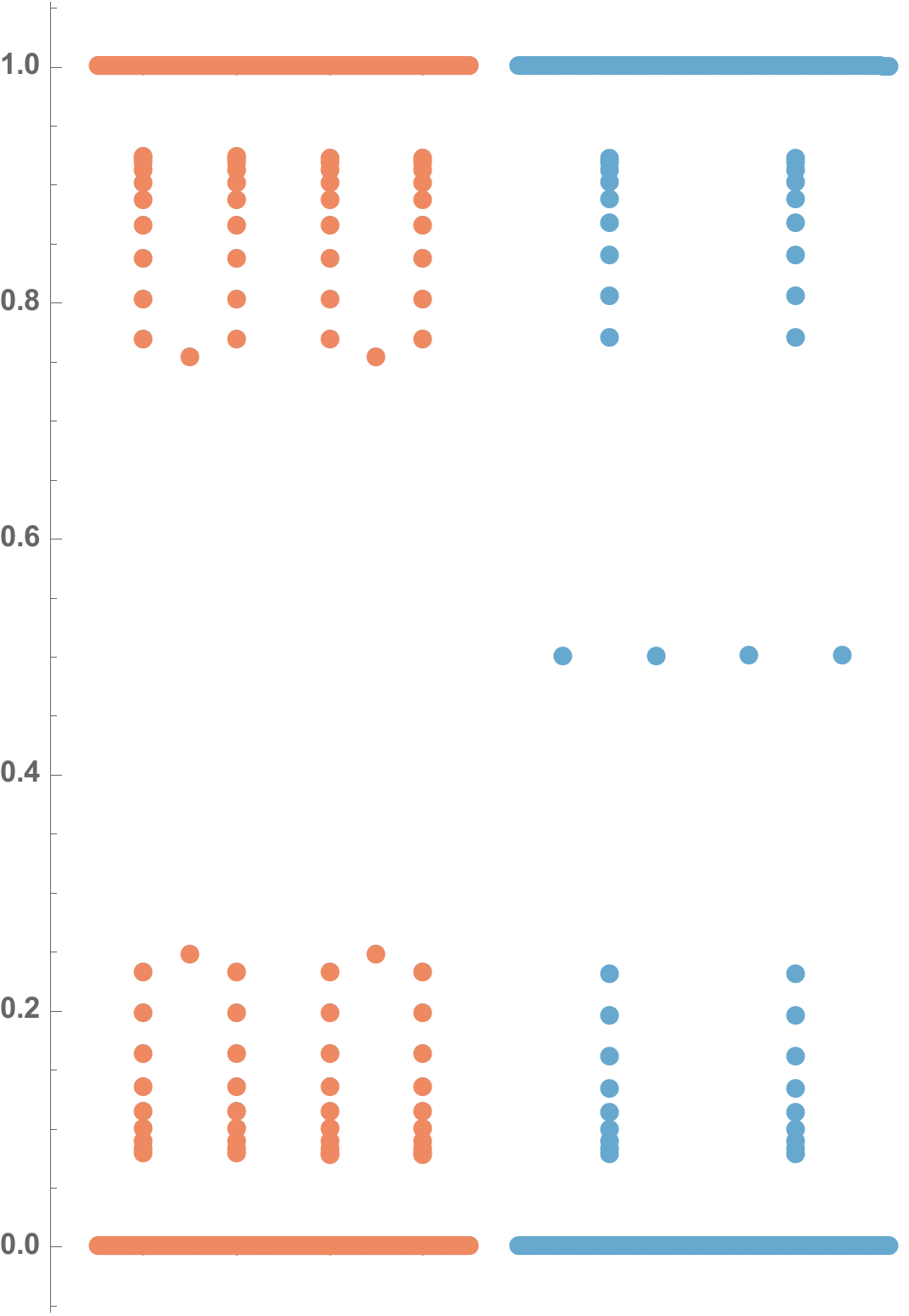}};
  \filldraw[fill=white, draw=white] (0,0) rectangle (0.2,5.5);
  \node at (0,5.28) {$\zeta$};
  \draw (0.2,0) -- (0.2,5.2);
  \node at (0,4.93) {1};
  \draw (0.15,4.93) -- (0.3,4.93);
  \node at (-0.1,2.61) {0.5};
  \draw (0.15,2.61) -- (0.3,2.61);
  \node at (0,0.28) {0};
  \draw (0.15,0.28) -- (0.3,0.28);
\end{tikzpicture}
\caption{Left: QTI model with two consecutive entanglement cuts displayed by red and blue lines. Right: Single particle entanglement spectra of the entanglement Hamiltonians obtained after performing the first (red) and the second (blue) cut. These spectra were computed for the quadrupole model with $t/\lambda=0.5$ on a periodic 20 by 20 lattice. The entanglement spectrum of the entanglement Hamiltonian obtained after making the first cut is gapped at half-filling, while spectrum of the 2nd nested entanglement Hamiltonian is gapless.}
\label{fig:quadrupole_spectra}
\end{figure}
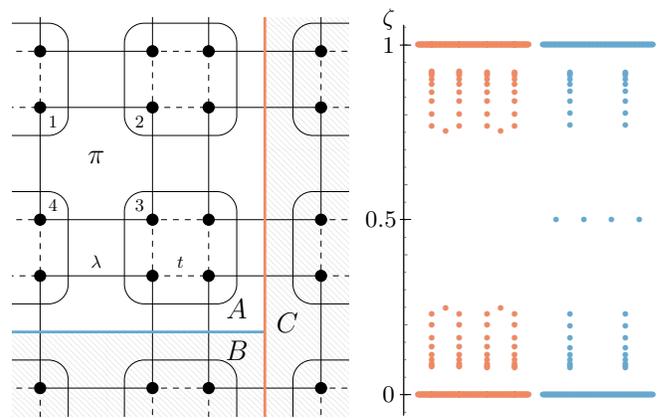

For the next step we consider a system described by the entanglement Hamiltonian $H^{E,1}_{AB}$ which acts only on the part of the Hilbert space corresponding to degrees of freedom located in the physical regions labeled by $A$ and $B$. In some simple cases we can readily identify this Hamiltonian as that of a gapped SPT state living on the entanglement cut. If a boundary SPT is not plainly apparent, then one can proceed by trying to characterize the ground state of the entanglement Hamiltonian using additional entanglement cut(s). For this procedure to be unambiguous we need the ground state of $H^{E,1}_{AB}$ to be unique. If the ground state is not unique (generically) then we would expect the either the entanglement ground state might spontaneously break some symmetry, or the system may already have some type of first-order SPT structure that leads to protected boundary/entanglement cut states that will obscure the identification of any higher order topology. For higher order phases we expect the physical edges to be gapped, and the (first-order) entanglement ground state to be unique. Therefore, the requirement of a unique ground state is precisely what one would expect for a HOSPT. Hence, to proceed we calculate the ground state of the entanglement Hamiltonian $H^{E,1}_{AB}$, and then make a second entanglement cut, e.g., the one labeled as $\circled{2}$ in Fig. \ref{fig:2d_cuts}. Tracing out the region $B$,  we find a reduced density matrix from which we can extract the 2nd nested entanglement Hamiltonian:
\begin{equation}
    \rho^{(2)}_{A}=\text{Tr}_B\rho^{(1)}_{AB}=\exp\left(-H^{E,2}_{A}\right).
\end{equation}
For some second-order HOSPTs we will find that the ground state of $H^{E,2}_A$ has protected degeneracy. Indeed, as we will see below the quadrupolar HOSPT has a doubly degenerate second-order entanglement spectrum (see Fig. \ref{fig:quadrupole_spectra}). Intuitively this arises because the initial entanglement Hamiltonian $H^{E,1}_A$, which is supported on an effectively 1D region, represents a first-order 1D SPT. The second entanglement cut then reveals the entanglement degeneracy corresponding to this SPT (expected for 1D SPTs from Ref. \onlinecite{ptbo}).  Hence, we propose that this procedure can serve as an important many-body characteristic for some HOSPTs. 

\section{Results}
\label{results}
In this section we will illustrate the usefulness of our procedure by  explicitly computing 2nd and 3rd order nested entanglement Hamiltonians for a series of models that are in quadrupolar and octupolar HOSPT phases respectively. We will consider a variety of bosonic and fermionic models, both free and interacting. In every case we will show that the first-order entanglement Hamiltonian bears the same spectral structure as the Hamiltonian of the physical boundary. In particular, we show that for the systems we study  the $(n-1)$-th nested entanglement Hamiltonian  $H^{E,n-1}$ represents a gapped, non-trivial SPT phase, in analogy to what is found in some HOSPT phases. Consequently, when further cut, such phases will yield an $n$-th order NEH that has a protected, doubly-degenerate eigenspectrum.

\subsection{QTI model}
We begin with the QTI model first introduced in Ref. \onlinecite{quadrupole}. It is a free-fermion, tight-binding model defined on a square lattice with four spinless fermion orbitals per unit cell. Hopping amplitudes are dimerized in both the $x$ and $y$ directions, and $\pi$-fluxes are threaded through each plaquette of the lattice. The real space structure of this model is shown in Fig. \ref{fig:quadrupole_spectra} where $\lambda$ and $t$ denote the amplitudes of the dimerized hoppings. For a system at half-filling this model is generically gapped, and can be tuned to two distinct HOSPT phases with the trivial insulating phase occurring for $|t|>|\lambda|,$ and the topologically non-trivial phase (exhibiting  fractional corner charges on open boundaries) for $|t|<|\lambda|$. 
The entanglement structure  of this model was thoroughly studied in Ref. \onlinecite{fukui18} (at least at the single particle level), where it was shown that the ground state of the entanglement Hamiltonian is gapped, and its unique ground state develops a Zak-Berry phase (polarization). This matches the expectations arising from considering the physical edge of this system, which has a quantized charge polarization in the presence of mirror symmetry, and the nested Wilson loop formalism\cite{quadrupole}. 

Our goal in this subsection is to apply our proposed method to the many-body ground state of this model to directly compute the first and second nested entanglement spectra for the QTI model.  As a first simple exercise, let us work out the ground state entanglement structure for this model in a zero correlation length limit where the intra-cell hopping amplitude $t$ vanishes. Here we will consider our model on an open infinite plane, although our discussion can be straightforwardly extended to finite or periodic lattices. The Hamiltonian in our case decomposes into a collection of disjointed $\pi$-flux plaquettes (four-site periodic chains) each at half-filling. They are described by the Hamiltonian (see site labels in Fig. \ref{fig:quadrupole_spectra}):
\begin{equation}
  H_{\square}= c^\dagger_1 c_2- c^\dagger_2 c_3+  c^\dagger_3 c_4+ c^\dagger_4 c_1,
  \label{eqn:square_ham}
\end{equation}
where we have enforced $C_4$ symmetry such that $\lambda_x=\lambda_y\equiv\lambda=1.$
Each of these clusters has a unique ground state: 
\begin{equation}
\begin{split}
    \ket{\psi^{(0)}_\square}=\frac1{2\sqrt{2}}\big(\ket{1100}+\ket{0110}&-\ket{1001}-\ket{0011}\big)\\
  &+\frac12\big(\ket{0101}-\ket{1010}\big),
  \end{split}
  \label{eqn:plq_gs}
\end{equation} where we have used the occupation basis.
Therefore, the ground state of the overall Hamiltonian is also unique as it is given by the tensor product of the ground states on each of the disjoint clusters:
\begin{equation}
    \ket{\Psi^{(0)}}=\bigotimes_i \ket{\psi^{(0)}_{\square,i}},
\end{equation}
where the index $i$ runs over all four-site clusters.

To carry out our procedure, let us first make an entanglement cut shown by the red line in Fig. \ref{fig:quadrupole_spectra}. The reduced density matrix is given by a tensor product of reduced density matrices for each individual cluster:
\begin{equation}
    \rho^{(1)}_{AB}=\text{Tr}_C\rho^{(0)}=\bigotimes_i\rho^{(1)}_{AB,i}
\end{equation}
where $\rho^{(0)}$ is the density matrix of the ground state $\ket{\Psi^{(0)}}$, $\rho^{(0)}_{AB,i}=\text{Tr}_C(\rho^{(0)}_{i}),$ and the index $i$ runs over all individual clusters. Let us denote by $\mathcal{C}$ the set of all individual clusters, and let $\mathcal{C}_{A}$ denote the subset of clusters that are completely contained in region $A,$ and similarly for other subsets of the lattice. Additionally, by $\mathcal{C}_{A\cap B}$ will denote clusters that overlap sites from both regions $A$ and $B$, but not any other region, i.e., clusters that lie directly on the boundary between $A$ and $B$. First, note that for a cluster $i$ that is completely contained in $AB$ (i.e., the region inside the first entanglement cut), we have $\rho^{(1)}_{AB,i}\equiv \rho^{(0)}_{i}$, since we are taking a partial trace over sites that are not included in cluster $i$. Second, for clusters that completely lie in $C$ (i.e., $i\in \mathcal{C}_{C}$) we have $\rho^{(1)}_{AB,i}\equiv 1$, as those groups of sites are traced out completely. Finally, there is a set of clusters that are split by the entanglement cut. The reduced density matrix for one of these clusters can be directly computed in the occupation number basis $\{\ket{00}_j,\ket{01}_j,\ket{10}_j,\ket{11}_j\}$ with $\ket{01}_j\equiv c^\dagger_{4,j}\ket{00}_j$, $\ket{10}_j\equiv c^\dagger_{1,j}\ket{00}_j$, where the operator $c^\dagger_{\alpha,j}$ creates a fermion on the $\alpha$-th site of the j-th cluster. The aforementioned reduced density matrix written in this basis is:
\begin{equation}
  \rho^{(1)}_{AB,j}=\frac18\left(\begin{matrix}
      1 & 0 & 0 & 0\\
      0 & 3 & -2\sqrt{2} & 0\\
      0 & -2\sqrt{2} & 3 & 0\\
      0 & 0 & 0 & 1
  \end{matrix}\right),
  \label{eqn:rho12num}
\end{equation}
where the index $j$ belongs to one of the plaquettes lying directly on the cut (i.e., $j\in \mathcal{C}_{AB\cap C}$).
The total reduced density matrix is then:
\begin{equation}
    \rho^{(1)}_{AB}=\bigotimes_{i\in \mathcal{C}_{AB}}\rho^{(0)}_{i}\bigotimes_{j\in \mathcal{C}_{AB\cap C}}\rho^{(1)}_{AB,j}.
\end{equation}

As we are working with disjoint clusters, the entanglement Hamiltonian (\ref{eqn:def_ent_ham}) constructed from this reduced density matrix is given by a tensor product of $H^{E,1}_{AB,i}=-\log(\rho^{(1)}_{AB,i})$ over all clusters $i$ that are included in the product above. The ground state of the total entanglement Hamiltonian is then simply a tensor product of the ground states for each individual $H^{E,1}_{AB,i}$. 
For clusters contained entirely in $AB$ we have:
\begin{equation}
    H^{E,1}_{AB,i}=-\log(\rho^{(0)}_{i})=-\log\left(\text{e}^{-H_{\square,i}}\right)=H_{\square,i}, 
\end{equation}
with $H_{\square,i}$ being exactly the Hamiltonian (\ref{eqn:square_ham}) acting on the $i$-th plaquette\footnote{We note that they are only \emph{exactly} the same here because we chose $\lambda=1$, otherwise they would differ by an overall scale factor}. 
Thus, the ground state of $H^{E,1}_{AB,i}$ for $i\in \mathcal{C}_{AB}$ is simply $\ket{\psi^{(0)}_{\square}}_i$ as in Eq. \ref{eqn:plq_gs}. 
Now, for clusters lying directly on the cut we have $H^{E,1}_{AB,j}=-\log(\rho^{(1)}_{AB,j})$ for the density matrix (\ref{eqn:rho12num}), therefore we can directly compute:
\begin{equation}
    H^{E,1}_{AB,j}=\left(\begin{matrix}
      \log(8) & 0 & 0 & 0\\
      0 & \log(8) & \log(3+2\sqrt{2}) & 0\\
      0 & \log(3+2\sqrt{2}) & \log(8) & 0\\
      0 & 0 & 0 & \log(8)
  \end{matrix}\right)
  \label{eqn:ent_ham_edge}
\end{equation}
for $j\in \mathcal{C}_{AB\cap C}$. 
The ground state of this entanglement Hamiltonian is unique and is given by:
\begin{equation}
    \ket{\psi^{(1)}}_j=\frac{1}{\sqrt{2}}\left(\ket{01}_j-\ket{10}_j\right).
    \label{eqn:dimer_gs_2}
\end{equation}

From this analysis we find that the total entanglement Hamiltonian is a tensor product of individual entanglement Hamiltonians acting on every cluster that remains in the physical region $AB$ after we have taken the partial trace over sites contained in region $C$:
\begin{equation}
    H^{E,1}_{AB}=\bigotimes_{i\in \mathcal{C}_{AB}}H_{\square,i}\bigotimes_{j\in \mathcal{C}_{AB\cap C}}H^{E,1}_{AB,j}.
    \label{eqn:1st_neh_qti}
\end{equation}
Its ground state takes the form:
\begin{equation}
    \ket{\Psi^{(1)}}=\bigotimes_{i\in \mathcal{C}_{AB}}\ket{\psi_{\square}}_i\bigotimes_{j\in \mathcal{C}_{AB\cap C}}\ket{\psi^{(1)}}_j,
    \label{eqn:he1_gs}
\end{equation}
where the index $j$ runs over clusters that lie on the edge of the entanglement Hamiltonian. Hence, the edge state of $H^{E,1}_{AB}$ is given by:
\begin{equation}
    \ket{\Psi^{(1)}_{edge}}=\bigotimes_{j\in \mathcal{C}_{AB\cap C}}\frac{1}{\sqrt{2}}\left(\ket{01}_j-\ket{10}_j\right),
\end{equation}
which is exactly the ground state of a dimerized Su-Schrieffer-Heeger (SSH)\cite{su1979} chain in the zero-correlation length limit. 
Therefore, the entanglement Hamiltonian computed for the ground state of the QTI model hosts a one-dimensional SPT phase at the entanglement edge (protected by the residual mirror symmetry on the edge).

To complete our analysis, let us now carry out the second-order (nested) entanglement cut as indicated by the blue line in Fig. \ref{fig:quadrupole_spectra}. From this we will compute the entanglement Hamiltonian $H^{E,2}_A$ by cutting the ground state $\ket{\Psi^{(1)}}$ of $H^{E,1}_{AB}$. 
The analysis here is exactly identical to the one laid out in the previous paragraphs. The nested entanglement Hamiltonian $H^{E,2}_A$ is given by the tensor product of entanglement Hamiltonians computed for individual clusters.  
For clusters entirely contained in $A$ we have $H^{E,2}_{A,i}\equiv H_{\square,i}$.
Similarly, the entanglement Hamiltonian computed for a cluster lying across the second cut, and entirely contained in $AB,$ is given by the matrix (\ref{eqn:ent_ham_edge}). 
Finally, for the single cluster lying at the intersection of both cuts, as shown in Fig \ref{fig:quadrupole_spectra}, we find that its entanglement Hamiltonian in the basis $\{\ket{0},\ket{1}\}$ is:
\begin{equation}
    H^{E,2}_{A,\mathcal{C}_{A\cap B\cap C}}=\left(\begin{matrix} \log (2) & 0\\
    0 & \log(2)
    \end{matrix}\right).
\end{equation}
Hence, the total second order Nested Entanglement Hamiltonian is:
\begin{equation}
    H^{E,2}_A=\left[\bigotimes_{i\in \mathcal{C}_A}H_{\square,i}\bigotimes_{j\in \mathcal{C}_{A\cap B}\cup \mathcal{C}_{A\cap C}}H^{E,1}_{A,j}\right]\otimes H^{E,2}_{A,\mathcal{C}_{A\cap B\cap C}},
    \label{eqn:h2_ent}
\end{equation}
where $H^{E,1}_{A,j}$ is the Hamiltonian (\ref{eqn:ent_ham_edge}) acting on the cluster $j$ that is lies across one entanglement cut, and $H^{E,2}_{A,\mathcal{C}_{A\cap B\cap C}}$ is the contribution to the second order NEH for the cluster that lies at the intersection of both cuts. 
To compute the eigenspectrum of $H^{E,2}_{A,G_{A\cap B\cap C}}$ we simply need to diagonalize each individual term in the tensor product.
We indeed find that the presence of the last term in (\ref{eqn:h2_ent}) renders the entire spectrum to be doubly-degenerate. 

As we noted above, the ground state of the first entanglement Hamiltonian contained an SPT phase localized at its edge. Indeed, we found the total entanglement ground state can be split in two parts:
\begin{equation}
    \ket{\Psi^{(1)}}=\ket{\Psi^{(1)}_{bulk}}\otimes\ket{\Psi^{(1)}_{edge}}.
\end{equation}
The second cut will split both the bulk and the edge. Similar to the first cut, the bulk part will reveal a gapped SPT on the cut. However, the cut  also splits the edge SPT phase $\ket{\Psi^{(1)}_{edge}}$ and thus, we naturally expect\cite{ptbo} to find that the entanglement spectrum computed for this cut has doubly degeneracy. Considering this model on a finite lattice with periodic boundary conditions, we will have to make two subsequent pairs of parallel entanglement cuts: first will cut out a cylinder to find an entanglement Hamiltonian that hosts two SPT phases localized on the opposite edges. The second pair of cuts splits both of these SPT phases at two points resulting in a total degeneracy of the entanglement Hamiltonian spectra to be $2^4=16$, as there will be exactly four clusters that lie on the intersection of two subsequent cuts. However, while the ground states of both bulk and boundary clusters contain one electron per two sites, each of the corner Hamiltonians admits both states $\ket{0}$ and $\ket{1}$ in its ground state. By projecting to the half-filled subspace of the  Hilbert space we find that the degeneracy of the spectrum computed for the second NEH reduces to 6, for the half-filled particle sector, as there will be two electrons that need to be distributed between the four corner clusters.

To verify that these features of the entanglement structure of the QTI model are still present even away from the zero-correlation length limit, we numerically study the single-particle entanglement spectrum\cite{Peschel04} of the QTI model on a $20\times 20$ (unit cell) periodic lattice with $t/\lambda=0.5,$ and find that the first-order entanglement spectrum for the red cut is gapped at half-filling (red spectrum in Fig. \ref{fig:quadrupole_spectra}). After performing the second cut, this time using the unique ground state of the entanglement Hamiltonian, we find the second nested entanglement spectrum to be gapless as shown by the blue spectrum in the right part of Fig. \ref{fig:quadrupole_spectra}. At half-filling, two of the four mid-gap states are filled rendering the overall degeneracy of the ground state to be exactly 6 as we predicted at the end of a previous paragraph. We note that this calculation is in agreement with the results of Ref. \onlinecite{fukui18} which argued that the entanglement edge of the QTI model has a non-trivial Berry phase/polarization. In Appendix \ref{app:fqti} we compare the QTI results with another free-fermion model having corner charges, but not a quadrupolar structure\cite{quadrupole2}. We indicate that our method fails to produce a result for this model because the first order entanglement cut has a ground state degeneracy, in analogy with what one would expect from the nested Wilson loop analysis\cite{quadrupole2}.

\subsection{Ring-Exchange model}
Let us now consider an interacting model that also has a quantized quadrupole moment: a quadrupolar ring-exchange model\cite{youburnellhughes2019,dubinkin2019dipole}. The ground state of this model is an example of a second order SPT phase protected by $C_4$ symmetry that is augmented by a set of (fine-tuned) $U(1)$ subsystem symmetries that impose charge conservation along every row and every column of the square lattice. Once again, the model is defined on the square lattice with four fermionic degrees of freedom per unit cell as shown on the left of Fig. \ref{fig:quadrupole_spectra}. Instead of constructing our model from dimerized hopping terms as in the QTI model, let us couple clusters of four neighboring sites by alternating \emph{ring-exchange} terms. These terms either act on four sites belonging to a single unit cell (represented by dashed squares) or on four sites belonging to four neighboring unit cells (represented by solid squares). The Hamiltonian can be written as:
\begin{equation}
    \begin{split}
    H=\lambda\sum_{\textbf{p}}\left|\swne\right\rangle\left\langle\nwse\right|_{\textbf{p}}+t\sum_{\textbf{s}} \left|\swne\right\rangle\left\langle\nwse\right|_{\textbf{s}}+h.c.
    \end{split}
    \label{eqn:Ham4pt}
\end{equation}
where $\left|\swne\right\rangle$ indicates a state of a four-site plaquette with an electron sitting at the upper right corner and one on the lower left corner. The index $\textbf{p}$ runs over all ring exchange terms acting on plaquettes (solid squares on the left side of Fig. \ref{fig:quadrupole_spectra}), while $\textbf{s}$ runs over ring exchange terms acting on-site (dashed squares on the left side of Fig. \ref{fig:quadrupole_spectra}). 

While we cannot explicitly solve for the ground state of this model for arbitrary values of $t$ and $\lambda$, we can analyze this model in the limit with $t=0.$ In this case the model simply decomposes into a collection of decoupled ring-exchange terms that we can consider individually in order to compute the entanglement spectrum of the ground state. The analysis we need to do in this simple case is exactly analogous to one performed in the previous subsection and the second order NEH has the same structure as  Eq. \ref{eqn:h2_ent}. Explicitly, let us first compute the entanglement Hamiltonian for a plaquette that is split between two regions (e.g. $AB$ and $C$). After making the first cut (shown in red in Fig. \ref{fig:quadrupole_spectra}), we find that the reduced density matrix for a single plaquette $j\in \mathcal{C}_{AB\cap C}$ takes the following form in the  $\{\ket{00}_j,\ket{01}_j,\ket{10}_j,\ket{11}_j\}$ occupation number basis:
\begin{equation}
    \rho^{(1)}_{AB,j}=\frac12\left(\begin{matrix}
      0 & 0 & 0 & 0\\
      0 & 1 & 0 & 0\\
      0 & 0 & 1 & 0\\
      0 & 0 & 0 & 0
  \end{matrix}\right).
\end{equation}
The corresponding entanglement Hamiltonian in the basis $\{\ket{01}_j,\ket{10}_j\}$ is:
\begin{equation}
    H^{E,1}_{AB,j}=\left(\begin{matrix}
      \log(2) & 0 \\
      0 & \log(2)
    \end{matrix}\right),
    \label{eqn:ent_re_h1}
\end{equation}
with the two states $\ket{00}_j$ and $\ket{11}_j$ having infinite energy, and thus effectively being projected out. 

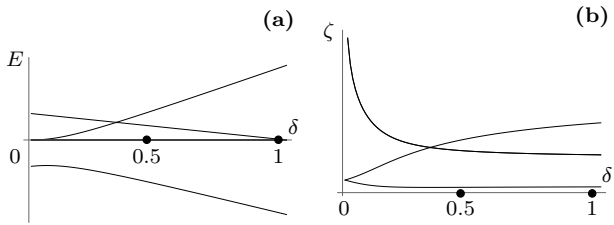
\begin{figure}
\begin{tikzpicture}
  \node[anchor=south west,inner sep=0] at (0,0) {\includegraphics[width=0.2\textwidth]{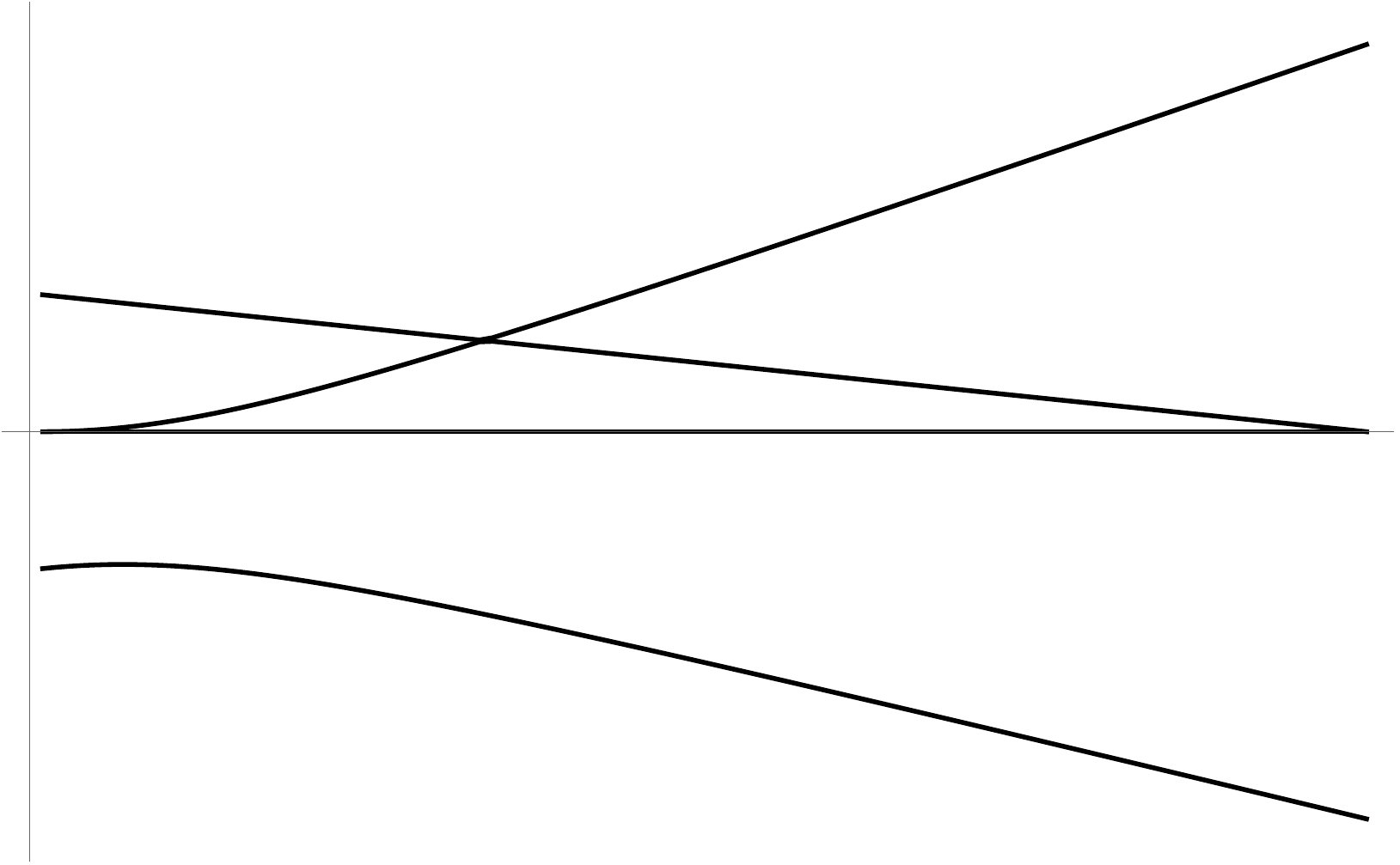}};
  \node at (3.4,2.7) {\footnotesize \textbf{(a)}};
  \node at (-0.1,2.2) {\footnotesize $E$};
  \node at (3.6,1.3) {\footnotesize $\delta$};
  \node at (3.4,0.9) {\footnotesize 1};
  \node at (1.65,0.9) {\footnotesize 0.5};
  \node at (-0.1,0.9) {\footnotesize 0};
  \node[draw,circle,inner sep=1pt,fill] at (3.4,1.12) {};
  \node[draw,circle,inner sep=1pt,fill] at (1.65,1.12) {};
\end{tikzpicture}
\begin{tikzpicture}
  \node[anchor=south west,inner sep=0] at (0,0) {\includegraphics[width=0.2\textwidth]{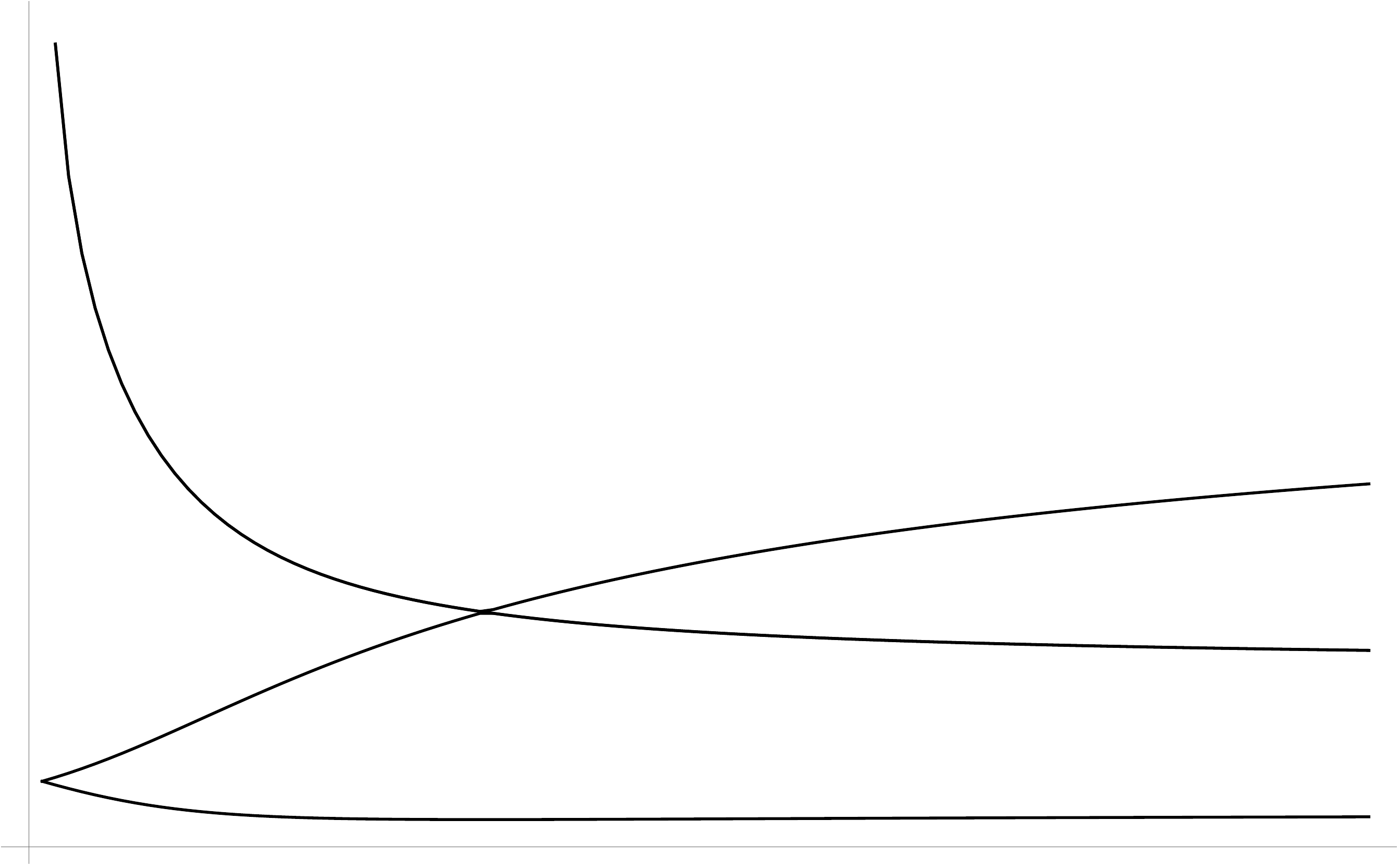}};
  \node at (3.4,2.4) {\footnotesize \textbf{(b)}};
  \node at (-0.1,2.2) {\footnotesize $\zeta$};
  \node at (3.6,0.3) {\footnotesize $\delta$};
  \node at (3.4,-0.15) {\footnotesize 1};
  \node at (1.65,-0.15) {\footnotesize 0.5};
  \node at (0.1,-0.15) {\footnotesize 0};
  \node[draw,circle,inner sep=1pt,fill] at (3.4,0.05) {};
  \node[draw,circle,inner sep=1pt,fill] at (1.65,0.05) {};
\end{tikzpicture}
\caption{Energy (a) and the ground state entanglement (b) spectra as we interpolate between the ring-exchange Hamiltonian and a free-fermion quadrupole topological insulator Hamiltonian. We can see that upon adding the $U(1)$ subsystem symmetry-breaking hopping terms the degeneracy of the entanglement ground states is immediately lifted. }
\label{fig:ring_exchange_spectra}
\end{figure}

Interestingly, unlike the fermion QTI model, this result indicates that the eigenspectrum of $H^{E,1}_{AB,j}$ is doubly degenerate. Hence, the eigenspectrum of the full first-order entanglement Hamiltonian $H^{E,1}_{AB}$ thus has a massive degeneracy of $2^N$ with the $N$ being the number of plaquettes lying directly on the cut. However, one can observe that this degeneracy in the entanglement spectrum is protected by the set of $U(1)$ subsystem symmetries that conserve charges on rows that end on the cut. These fine-tuned symmetries may be broken while preserving the $C_4$ and mirror symmetries that protect the bulk quadrupole phase. 
We can do so by adding a small perturbation of strength $\varepsilon$, in the form of (quadratic) hopping terms, to each inter-cell link of the lattice that runs parallel to the cut. This modifies the entanglement Hamiltonian of a single cluster on the cut to the following form:
\begin{equation}
    \tilde{H}^{E,1}_{AB,j}=\left(\begin{matrix}
      \log(2) & \varepsilon \\
      \varepsilon & \log(2)
  \end{matrix}\right)
  \label{eqn:ent_re_ht_pert}
\end{equation}
having eigenvalues $\log(2)\pm\varepsilon$. Thus, this perturbation immediately breaks the degeneracy (see Fig. \ref{fig:ring_exchange_spectra} (b)), and yields a unique ground state:
\begin{equation}
    \ket{\psi^{(1)}}_j=\frac{1}{\sqrt{2}}\left(\ket{01}_j-\ket{10}_j\right),
    \label{eqn:dimer_gs}
\end{equation} for the first order entanglement Hamiltonian of the cluster $j.$ 
This is exactly the same ground state we found at the entanglement edge of the QTI model.
We remark that, when considering an actual physical edge which is realized by turning off a line of ring-exchange plaquette terms, one similarly finds the massive degeneracy in the spectrum of the physical Hamiltonian due to the presence of the edge modes. However, at the physical edge, such edge modes can be gapped out without breaking $U(1)$ subsystem symmetries: one simply has to add intra-cell hopping terms between the dangling fermionic sites. 
However, when the entanglement edge is considered, adding intra-cell hopping terms does not lift the degeneracy, at least perturbatively. To see this, let us consider a pair of edge clusters with the single intra-cell term that connects them. 
Such a hopping term necessarily maps the ground state subspace of two clusters $\{\ket{01}_1,\ket{10}_1\}\otimes\{\ket{01}_2,\ket{10}_2\}$ to $\{\ket{00}_1,\ket{11}_1\}\otimes\{\ket{00}_2,\ket{11}_2\}$, which is the part of the Hilbert space that is projected out by the entanglement Hamiltonian (\ref{eqn:ent_re_h1}).

With the inter-cell hopping terms turned on, the entanglement Hamiltonian computed after making the first cut is given by:
\begin{equation}
    H^{E,1}_{AB}=\bigotimes_{i\in \mathcal{C}_{AB}}H_{\square,i}\bigotimes_{j\in \mathcal{C}_{AB\cap C}}\tilde{H}^{E,1}_{AB,j}.
\end{equation}
with $H_{\square,i}=\left|\swne\right\rangle\left\langle\nwse\right|_j+h.c.,$ and $H^{E,1}_{AB,j}$ given by Eq. (\ref{eqn:ent_re_ht_pert}).
Since this entanglement Hamiltonian has a unique ground state, we can move on to perform the nested entanglement cut and compute the second order entanglement spectrum. 
Following the logic from the previous subsection that dealt with the QTI model, we can write the second NEH in the following form:
\begin{equation}
    H^{E,2}_A=\bigotimes_{i\in \mathcal{C}_A}H_{\square,i}\bigotimes_{j\in \mathcal{C}_{A\cap B}\cup \mathcal{C}_{A\cap C}}\tilde{H}^{E,1}_{A,j}\otimes H^{E,2}_{A,\mathcal{C}_{A\cap B\cap C}}
    \label{eqn:h2_ent_re}
\end{equation}
where we perturbatively added both vertical and horizontal inter-cell hopping terms to break all of the $U(1)$ subsystem symmetries.
Finally, we need to compute the second NEH for the dimer lying at the intersection of both cuts. We find the reduced density matrix for this dimer to be  $\rho^{(2)}_{A,\mathcal{C}_{A\cap B\cap C}}=\frac12\mathbb{I}_{2\times 2},$ and thus the second entanglement Hamiltonian $H^{E,2}_{A,\mathcal{C}_{A\cap B\cap C}}=\log(2)\mathbb{I}_{2\times 2}$ and the full second NEH for the ring-exchange model thus has a doubly-degenerate eigenspectrum.

To gain further intuition about this model we also showed that the ground state of the ring-exchange model in the zero-correlation length limit is adiabatically connected to the ground state of the QTI model, while preserving, e.g., $C_4$ and $U(1)$ charge conservation symmetries. Explicitly we study the following interpolation Hamiltonian for a single plaquette:
\begin{equation}
    H_\square(\delta)=\delta H^{QTI}_{\square} + (1-\delta)H^{RE}_{\square}
\end{equation}
where $H^{QTI}_\square$ is the Hamiltonian for an elementary plaquette of the QTI model, and $H^{RE}_\square$ is the ring-exchange Hamiltonian for the same plaquette.
In Fig. \ref{fig:ring_exchange_spectra}a,b we show the energy spectrum of $H_\square(\delta)$ and entanglement spectrum as we increase the parameter $\delta$ from 0 to 1. The entanglement cut traces out half of the plaqeutte sites in the ground state of $H_\square(\delta)$. We hence see that these models are adiabatically connected (while preserving the protective symmetries), and this confirms our result that the nested entanglement spectra should show the same features as long as the subsystem symmetries are broken.

\subsection{Bosonic XY model}
The next model we consider is  a second order HOSPT spin model\cite{Dubinkin18}. It is a direct bosonic counterpart of the QTI model considered above. 
This model is defined on the same lattice as the QTI, where we simply replace fermionic orbitals  with spin-1/2 degrees of freedom coupled via antiferromagnetic XY interaction terms instead of fermionic hopping terms. The lattice structure exactly matches the one shown in Fig. \ref{fig:quadrupole_spectra}.
Once again, working in a zero correlation length limit we can compute the second NEH for disjoint spin clusters. Each individual cluster has the following Hamiltonian:
\begin{equation}
    H_{\square}=\sum_{a=x,y}\left(\lambda_x \sigma^a_1 \sigma^a_2+\lambda_y \sigma^a_2 \sigma^a_3 + \lambda_x \sigma^a_3 \sigma^a_4+\lambda_y \sigma^a_4 \sigma^a_1\right).
\end{equation}
This calculation is very similar to the derivation of $H^{E,2}_A$ for the QTI model in the zero correlation length limit, and so we present the technical details in Appendix \ref{app:ent_edge_2d_heis} and just summarize the results here. 

For the XY model we find that the first and second order NEHs exhibit a similar overall structure to the corresponding entanglement Hamiltonians of the QTI model given by Eqs. (\ref{eqn:1st_neh_qti}) and (\ref{eqn:h2_ent}). Once again, the bulk structure of both NEHs is indistinguishable from the bulk of the original physical Hamiltonian. The interesting pieces of the reduced density matrix are located at the entanglement edge, for which we need to focus our attention on spin clusters located either on one of the cuts, or at the intersection(s) of both entanglement cuts as depicted in Fig. \ref{fig:quadrupole_spectra}. We find the entanglement Hamiltonian for any cluster lying on one of the cuts to have a unique ground state, leading to a unique ground state for the full first order NEH. The part of the entanglement Hamiltonian that is localized at the entanglement edge represents the Hamiltonian of an inter-cell dimerized spin-1/2 chain, which has an SPT ground state protected by the $\mathbb{Z}_2\times\mathbb{Z}_2$ symmetry group generated by: 
\begin{equation}
P_1=\prod_{\textbf{i}}\sigma^x_{\textbf{i}},\quad P_2=\prod_{\textbf{i}}\sigma^y_{\textbf{i}}
\label{eqn:XY_symmetry}
\end{equation}
where the index $\textbf{i}$ runs over each spin of the edge chain. Performing the second (nested) cut we find that the second order NEH computed for the cluster lying at the intersection of both cuts has a doubly-degenerate eigenspectrum leading to a doubly-degenerate spectrum of the full second order NEH.

Once we tune the couplings away from the zero correlation length limit, this model is no longer exactly solvable and doing the exact diagonalization is intractable for large size 2D systems. However, we can still provide a perturbative argument showing that the claimed features of the entanglement structure of the ground state are still present even away from the zero correlation length limit.

\begin{figure}
\includegraphics[width=0.5\textwidth]{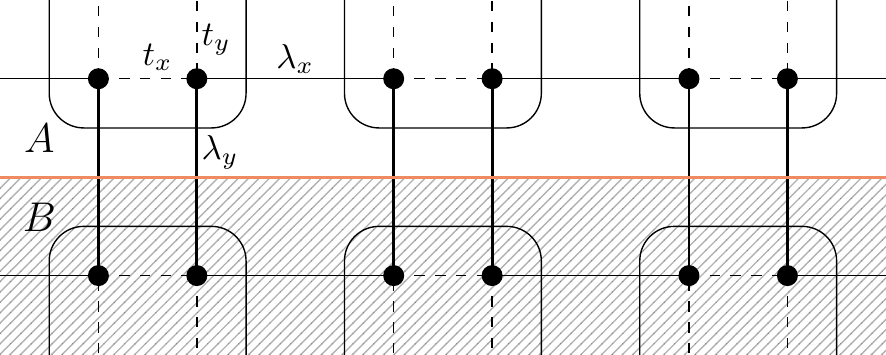}
\caption{Partition of a single line of plaquettes in two halves.}
\label{fig:line_x}
\end{figure}

The goal of the next two paragraphs will be to demonstrate that the edge of the first entanglement Hamiltonian $H_{A}^{E,1}$ stays in the same SPT phase even when the perturbative corrections are taken into account. Let us tune  the intra-cell coupling value $t_x=t_y=t$ to be non-zero.
Then, consider an entanglement cut made along the $\hat{x}$-axis that splits our system in two halves $A$ and $B,$ as shown in Fig \ref{fig:line_x}. To compute the entanglement Hamiltonian $H^{E,1}_A$ up to first order corrections it is sufficient to consider the set of terms around the cut:
\begin{equation}
  H^{cut}=\lambda_x H^X+\lambda_y H^Y+tH_A+tH_B   \label{Ham_plaq_chain}
\end{equation}
where $H^X$ and $H^Y$ are inter-cell XY couplings across the cut, and $H_A$ and $H_B$ describe sets of intra-cell XY couplings on opposing sides of the cut, as shown in Fig. \ref{fig:line_x}.  

Now, we are going to imbalance the value of the inter-cell couplings in different directions, so that $\lambda_y\gg \lambda_x\gg t$.
Crucially, the off-set in strengths of $\lambda_x$ and $\lambda_y$ leaves the model in the same topological phase since tuning $\lambda_y\gg\lambda_y$ preserves both the bulk and the edge gaps, provided that $|\lambda_y|>|t|$. Performing an entanglement cut, and treating both $H^X$ terms and intra-cell couplings $tH_A+tH_B$ perturbatively with the small parameter $t=\varepsilon,$ we find that the first order perturbative correction to the entanglement Hamiltonian $H^{E,1}_A$ retains the structure of the dimerized spin-1/2 chain (see Appendix \ref{app:ent_edge_2d_heis}) for details.
 This implies that a non-trivial edge SPT survives even after we perturbatively tune away from the zero-correlation length limit. Hence, making a second order (nested) cut that splits this 1D SPT phase will still result in a second order NEH with a doubly degenerate eigenspectrum.

\textit{3D XY model.} It is possible to extend the 2D XY model to construct an octupolar HOSPT phase in 3D. 
Once again, we consider a direct counterpart of the 3D fermionic octupole model\cite{quadrupole,quadrupole2} where we replace fermions with spins and fermionic hopping terms with XY interactions. 
This model has gapped surfaces and hinges but has gapless corner modes (more precisely, fractional charges).
Therefore, we expect the third order nested entanglement Hamiltonian to have a doubly degenerate spectrum, while the first and second order NEHs will have a unique ground state. 
In a zero-correlation length limit, we can directly work out the entanglement structure of the ground state.
Having intra-cell coupling values $t$ set to zero, this model decomposes into a collection of disjoint cubes, and we can compute the entanglement Hamiltonian separately for each  cube.
For a single cut we find that the entanglement Hamiltonian computed for each of the dissected cubes represents a Hamiltonian for a single XY plaquette, having exactly the same structure as the 2D version of this model. Therefore, the entanglement Hamiltonian hosts a second order HOSPT model at the entanglement cut which leads us to conclude that the third order NEH $H_C^{E,3}$ for the cube $C$ located at the intersection of three consecutive cuts is proportional to the identity matrix and yields a doubly-degenerate eigenspectrum, while the $H_C^{E}$, computed for any other cube sliced by either 0, 1 or 2 cuts, has a unique ground state. 
Therefore, the full third order NEH $H^{E,3}$ has a doubly degenerate entanglement spectrum while both $H^{E,1}$ and $H^{E,2}$ have a unique ground state. This indicates that the ground state of this model is in an octupolar HOSPT phase. We can also turn on the intra-cell couplings $t$ and perform the same perturbative analysis we did for the 2D XY model to show that the entanglement Hamiltonian structure of the octupolar ground state is perturbatively stable away from the zero-correlation length limit.

\subsection{$\mathbb{Z}_2\times\mathbb{Z}_2$ model.}
Now, let us see how our characterization works for another bosonic quadrupolar second order SPT. Consider the $\mathbb{Z}_2\times\mathbb{Z}_2$ HOSPT model introduced in Ref. \onlinecite{Dubinkin18}. This is a simple 2D model consisting of a collection of pairs of $\mathbb{Z}_2\times\mathbb{Z}_2$ chains\cite{Suzuki1971,Raussendorf2001,Kopp2005} each described by a bosonic Hamiltonian
\begin{equation}
  H_{\mathbb{Z}_2\times\mathbb{Z}_2}=-\sum_{i=1}^{N-1}\left(Z^a_{i}X^b_iZ^a_{i+1}+Z^b_iX^a_{i+1}Z^b_{i+1}\right)
  \label{eqn:z2z2_ham}
\end{equation}
where pairs of chains (which we denote as $A$ and $B$) in neighboring unit cells are coupled by a set of additional `vertical' terms:
\begin{equation}
  \begin{split}
  &H_{V}=-\sum^{N}_{i=1} (Z^a_{i,A}Z^a_{i,B}+Z^b_{i,A}Z^b_{i,B}\\
  &+Z^a_{i,A}X^b_{i,A}Z^a_{i,B}X^b_{i,B}+X^a_{i,A}Z^b_{i,A}X^a_{i,B}Z^b_{i,B}),
  \label{eqn:vertical_terms}
\end{split}
\end{equation}
such that, on an open lattice, both horizontal edges end up carrying a single dangling $\mathbb{Z}_2\times\mathbb{Z}_2$ SPT chain\cite{Dubinkin18}. 
On a periodic lattice, this model represents a set of coupled pairs of periodic $\mathbb{Z}_2\times\mathbb{Z}_2$ SPT chains, where every chain has a partner. As was shown in Ref. \onlinecite{Dubinkin18}, this system is invariant with respect to $C_4$ rotations which is important for protecting corner modes. 

To simplify our derivations, we can pick the first entanglement cut to run along the $\mathbb{Z}_2\times\mathbb{Z}_2$ chains. 
Since chains are coupled only in pairs, to obtain the edge part of the entanglement Hamiltonian, all we need to consider is one such pair where we denote the $\mathbb{Z}_2\times\mathbb{Z}_2$ chain that lies above the cut by $A$ and the one below it by $B$. The full Hamiltonian we need to consider consists of three blocks:
\begin{equation}
  H=H_A+H_B+H_V
  \label{eqn:pair_of_z2z2}
\end{equation}
where $H_A$ and $H_B$ are both given by (\ref{eqn:z2z2_ham}) acting on chains $A$ and $B$ respectively.
To obtain the entanglement Hamiltonian, we are going to treat $H_A$ and $H_B$ perturbatively relative to the $H_V$. Similarly to the XY model case, this offset in strength of horizontal and vertical couplings leaves the ground state in the same phase. For such a system we can employ the result obtained in Ref. \onlinecite{Peschel2011}, which showed that if the following three conditions are true, then the entanglement Hamiltonian will be proportional to the Hamiltonian of the upper chain, i.e., $H^{E,1}_A\propto H_A$. 
These conditions are: 
\begin{itemize}
    \item $H_A$ and $H_B$ only couple the ground state $\ket{\psi_0}$  of $H_{V}$ to excited states $\ket{\psi_k}$ with the same gap $\Delta=E_k-E_0$.
    \item Both $H_A$ and $H_B$ have the same matrix elements in the eigenbasis of $H_V$: $\bra{\psi_k}H_A\ket{\psi_0}=\bra{\psi_k}H_B\ket{\psi_0}$. 
    \item The reduced density matrix for the unperturbed ground state is proportional to the identity matrix $\rho^0_A=\text{Tr}_B\ket{\psi_0}\bra{\psi_0}\propto \mathbb{I}$.
\end{itemize}
As $H_V$ represents a collection of terms acting on disjointed clusters of four spins, we can directly verify, through exact diagonalization, that all three conditions are satisfied for our model. 
The entanglement Hamiltonian for this cut is therefore proportional to the Hamiltonian of a single $\mathbb{Z}_2\times\mathbb{Z}_2$ SPT chain, which means that the ground state of the entanglement Hamiltonian is the well known $\mathbb{Z}_2\times\mathbb{Z}_2$ SPT phase which is exactly the same phase that appears on the one-dimensional edge of the two-dimensional $\mathbb{Z}_2\times\mathbb{Z}_2$ HOSPT model. 
Consequently, if we make an entanglement cut in the entanglement Hamiltonian we will find the eigenspectrum of the second nested entanglement Hamiltonian to be doubly-degenerate, indicating a quadrupole-like HOSPT.

\subsection{Topological Plaquette Paramagnet (TPP) model.}
\begin{figure}
  \includegraphics[width=0.5\textwidth]{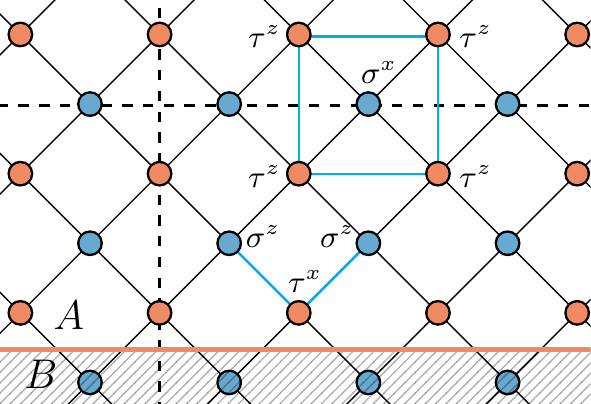}
  \caption{Topological Plaquette Paramagnet lattice model. Spin-1/2 degrees of freedom $\tau$ and $\sigma$ are associated with red and blue sites respectively. Two dashed lines indicate two sets of spins on which two different $\mathbb{Z}_2$ subsystem symmetry operators act. Light blue frame represents one of the stabilizers of the model, while the blue wedge depicts one of the terms that we have to add to break the subsystem symmetry and gap out the entanglement edge.}
\label{fig:2d_tpp}
\end{figure}

Let us consider another $\mathbb{Z}_2\times\mathbb{Z}_2$ model that additionally has a set of (fine-tuned, for our purposes) subsystem symmetries as described in Ref. \onlinecite{You18}. This is a bosonic model defined on a square lattice with spin degrees of freedom $\sigma$ and $\tau$ living on two different square sublattices governed by the following commuting-projector Hamiltonian:
\begin{equation}  H_{TPP}=-\sum_{i\in a}\tau^x_i\prod_{j\in P_i}\sigma^z_j-\sum_{i\in b}\sigma^x_i\prod_{j\in P_i}\tau^z_j,
  \label{eqn:Ham_z2z2_2}
\end{equation}
where $a$ and $b$ denote the red and blue sublattices respectively (arranged as depicted in Fig. \ref{fig:2d_tpp}).
 All individual terms in the Hamiltonian commute with every other term, and every individual term squares to 1. There is exactly one spin degree of freedom per stabilizer operator, and since there are no additional non-local constraints, this model is exactly solvable and has a unique ground state on a periodic lattice. As noted, in addition to the global $\mathbb{Z}_2\times\mathbb{Z}_2$ symmetry generated by the operators 
\begin{equation}
    \prod_{i\in \text{bulk}}\sigma^x_i\,\,\,\text{ and }\prod_{i\in \text{bulk}}\tau^x_i,
\end{equation}
this model has a set of subsystem $\mathbb{Z}_2$ symmetries generated by: 
\begin{equation}
    \prod_{i\in \text{row/column}}\sigma^x_i\,\,\,\text{ and }\prod_{i\in \text{row/column}}\tau^x_i,    
\end{equation}
where the product now runs over the spins located on the same row or column of the lattice. 

We can gain an understanding of the ground state of this model working in the $\tau^z$ and $\sigma^x$ basis. There, the second set of terms in Eq. (\ref{eqn:Ham_z2z2_2}) requires that for any plaquette $P_i$ of $\tau$ spins centered around a single $\sigma$ spin, we must have $\prod_{i\in P_i}\tau^z_j$ equal to $+1$ or $-1$ when the spin in the center of $P_i$ is in the state with $\sigma^x_i$ equal to $+1$ or $-1$ respectively. Thus, one can see that any configuration of $\tau^z_i=\pm 1$ across the lattice is acceptable in the ground state as long as any \emph{corners} of domain walls between $\tau^z$ spin-`up' and spin-`down' regions are decorated by $\sigma^x=-1$ states. The first set of operators in the Hamiltonian (\ref{eqn:Ham_z2z2_2}) simply maps between different configurations of corner-decorated domains allowed by the second set of terms, as it flips the value of a single $\tau^z$ spin operator along with the values of four $\sigma^x$ operators around it. Let us introduce a set of all possible products of $\tau^x_i\prod_{j\in P_i}\sigma^z_j$ operators, where $i\in a$. These operators form an Abelian group which we denote as $G$. Starting from the state $\ket{0}$ which has all $\tau^z=+1$ and all $\sigma^x=+1$ we can write down the ground state of Eq.  (\ref{eqn:Ham_z2z2_2}) as:
\begin{equation}
    \ket{\Psi^{(0)}}=\frac{1}{|G|}\sum_{g\in G}g\ket{0},
\end{equation}
which is simply the equal-weight superposition of all possible configurations of $\tau^z=\pm 1$ with corner-decorated domain walls. $|G|$ is the rank of the group $G$. 

Let us now make an entanglement cut running along one of the diagonals of the lattice that splits the system into two regions $A$ and $B$, as shown in Fig \ref{fig:2d_tpp}. 
To derive the entanglement Hamiltonian we employ the method described in Ref. \onlinecite{Flammia09}. There
it was shown that to derive the reduced density matrix  $\rho_A^{(1)}$ for this type of commuting-projector model we need to focus on the contributions that are coming from only the subspace $\mathcal{H}_{\partial A}\otimes\mathcal{H}_{\partial B},$ of the total Hilbert space $\mathcal{H},$ i.e., the subspace located near the cut. 

In our case the Hilbert subspace of interest corresponds to two pairs of lines of $\tau$ and $\sigma$ spins located above and below and cut. 
We define a quotient group $G_{AB}=G/\left[G_A\times G_B\right]$ where $G_A$ and $G_B$ are subgroups of $G$ having elements which have support on only region $A$ or $B$ respectively. 
Any element of $G_{AB}$ corresponds to a set of operators which have support simultaneously on both $A$ and $B$ regions. 
Clearly, from each coset in $G_{AB}$ we can pick a single representative element $g$ that has support on only the spins living near the boundary, i.e., that acts non-trivially on only the $\mathcal{H}_{\partial A}\otimes\mathcal{H}_{\partial B}$ subspace of $\mathcal{H}$. Furthermore, every such representative operator $g\in G_{AB}$ can be decomposed as a product of two operators acting on spins located on different sides of the cut: $g=g_A\otimes g_B$. As was shown in Ref. \onlinecite{Flammia09}, the reduced density matrix is then given by:
\begin{equation}
  \rho^{(1)}_A=\frac{1}{|G_{AB}|}\sum_{g\in G_{AB}}g_A\ket{0}_A\bra{0}_A g_A
  \label{eqn:RDM_TPP}
\end{equation} Where $\ket{0}_A$ is the restriction of $\ket{0}$ to the $A$ region of the lattice.

Applying this prescription to the TPP model, we find that the resulting entanglement Hamiltonian at edge is represented by the sum of $g_{A}$ terms, from which we see:
\begin{equation}
    H^{E,1}_A=\sum_{i\in a\in \partial A} \sigma_{i-1}^z\tau^x_{i}\sigma^z_{i+1},
\end{equation}
where the index $i$ runs over the sites of sublattice $a$ lined up along the entanglement cut depicted in Fig. \ref{fig:2d_tpp} on the region $A$ side. 
This Hamiltonian $H_A^{E,1}$ has a $2^N$ degenerate ground state subspace, where $N$ is the number of $\tau$ spins near the cut.
Evidently, the entanglement edge of the TPP model is gapless, just as the physical edge of this model. One can show, that gapless modes at the entanglement edge have exactly the same nature as the gapless modes that would appear at the edge if we were making a physical cut instead. 

Importantly, Ref. \onlinecite{You18} showed that the physical edge's gapless states are protected by a set of $\mathbb{Z}_2$ subsystem symmetries acting along the columns of $\tau^x$ spins.  
These modes can be gapped by introducing a set of subsystem symmetry-breaking terms at the physical edge. 
Let us add these same terms to our Hamiltonian and study their effects on the structure of the ground state. Consider the following perturbation to the original Hamiltonian (\ref{eqn:Ham_z2z2_2}):
\begin{equation}
  V=-\sum_{i\in b\in \partial A}\tau^z_{i-1}\sigma^x_i\tau^z_{i+1}
  \label{eqn:ssbreaking_term}
\end{equation}
where the index $i$ runs over sites located at the $A$-region's edge of the cut (for this model the terms added in the bulk away from the cut do not have an effect on the entanglement properties of the ground state). Let us now study the ground state $\ket{\tilde{\Psi}^{(0)}}$ of the perturbed Hamiltonian 
\begin{equation}
    \tilde{H}_{TPP}=H_{TPP}+\varepsilon V,
\end{equation}
where the parameter $\varepsilon$ is taken to be small. As we show in Appendix \ref{app:2d_tpp}, first-order perturbative correction to the entanglement Hamiltonian is simply proportional to $V$. 
The resulting collection of terms appearing at the entanglement edge forms a commuting-projector Hamiltonian with a unique ground state:
\begin{equation}
    \tilde{H}^{E,1}_A=H^{E,1}_A+\delta V,
\end{equation}
where $\delta$ is the proportionality coefficient.

We have now shown that the modified entanglement Hamiltonian near the cut is in fact the Hamiltonian of a one-dimensional $\mathbb{Z}_2\times\mathbb{Z}_2$ chain, where different species of stabilizers have different coefficients. However the offset in the couplings' strengths doesn't change the structure of the ground state, hence it has the exact same properties the same as the ground state of the well-known $\mathbb{Z}_2\times\mathbb{Z}_2$ SPT chain (that we also discussed above). This is exactly the same Hamiltonian one would obtain at the physical edge of this model after breaking the subsystem symmetries. Moreover, as the entanglement Hamiltonian computed for the TPP model having broken subsystem symmetries hosts a non-trivial SPT phase near the entanglement cut, the 2nd NEH $H^{E,2}$ then necessarily has a two-fold degenerate eigenspectrum.
This allows us to place the TPP model with broken subsystem symmetries into the same category as the $\mathbb{Z}_2\times\mathbb{Z}_2$ HOSPT model discussed in the previous section.

\textit{3D TPP model.} We can also see how our approach works for the 3D version of the TPP model\cite{You18}. There we find that after making the first cut, the entanglement edge Hamiltonian supports gapless modes protected by a set of subsystem symmetries. Breaking these symmetries by introducing a set of boundary terms to gap out the physical edge of the 3D TPP model\cite{You18}, we obtain a 2D TPP model at the entanglement edge (see Appendix \ref{app:ent_edge_2d_heis}). This is the same phase one expects to see at the physical edge of this model after breaking subsystem symmetries.
Crucially, after breaking subsystem symmetries, we find that the entanglement edge Hamiltonian is exactly the Hamiltonian of the 2D TPP model (\ref{eqn:Ham_z2z2_2}) -- a second order HOSPT system.
From this point we can see that making two subsequent nested entanglement cuts, while breaking appropriate subsystem symmetries as we did while considering the 2D model, yields a Hamiltonian $H^{E,3}$ that has a doubly degenerate eigenspectrum. Therefore, the 3D TPP model with broken subsystem symmetries provides us with another example of an octupolar-like HOSPT phase.

\subsection{3D CZX model}
\begin{figure}
\includegraphics[width=0.5\textwidth]{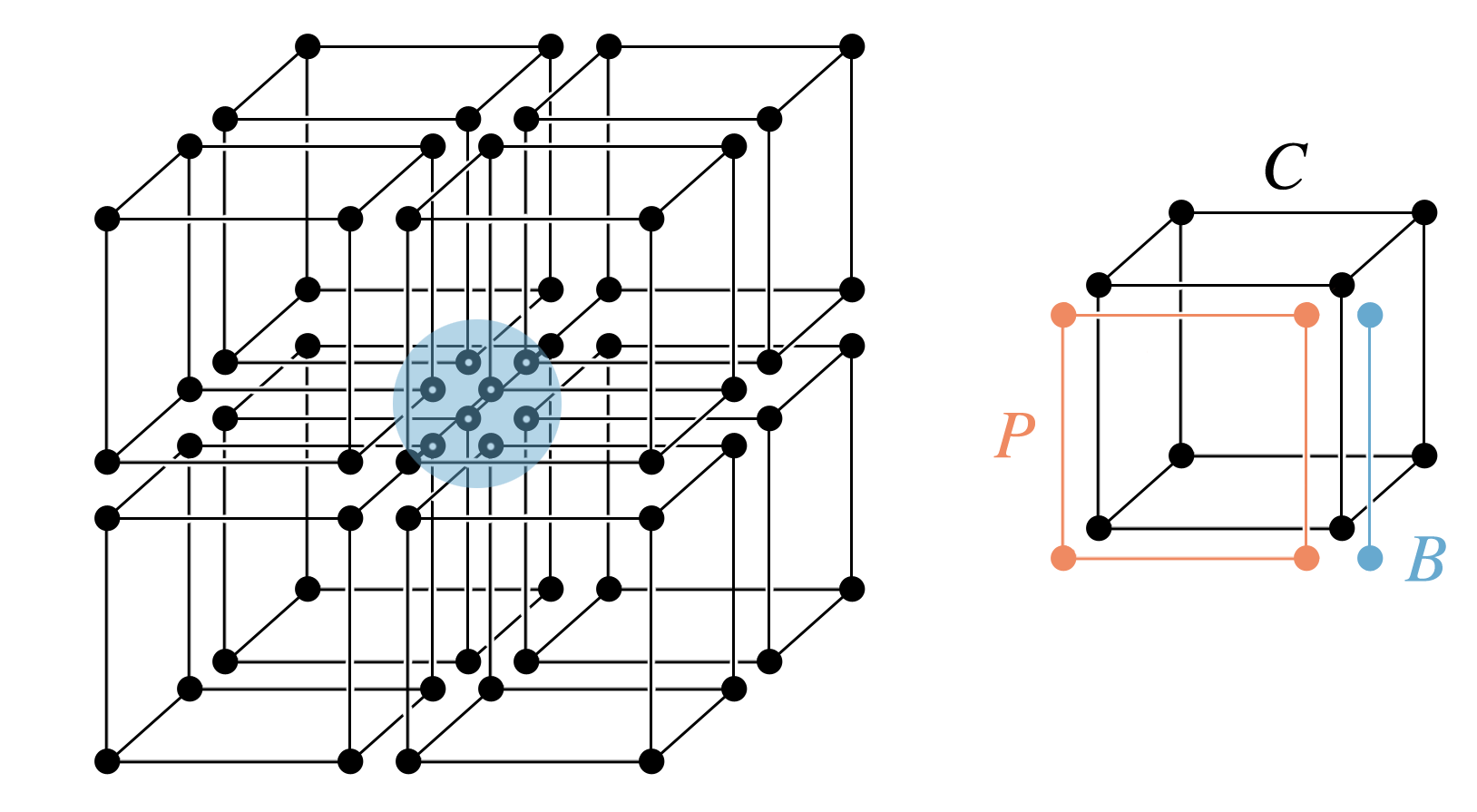}
\caption{3D CZX model. With each cube $C$ of the lattice we associate a term in the Hamiltonian that is a product of the cubic term $H_C$, 6 boundary plaquette terms $H_P$ (one of which is shown on the figure in red), and 12 boundary bond terms $H_B$ (on of which is shown on the figure in blue color). These terms are explicitly written in Eq. \ref{eqn:3dczx_ham}.}
\label{fig:3dczx}
\end{figure}
The final model we consider represents a three-dimensional HOSPT system with additional subsystem symmetries\cite{YouDevakul18}. It is defined on a 3D cubic lattice with eight spin degrees of freedom per unit cell as shown in Fig. \ref{fig:3dczx}. The interaction terms in the Hamiltonian involve eight spin degrees of freedom from the sites at the corners of each cube $C$, 24 spins from 6 plaquettes $P$ neighboring the faces of the cube, and 24 spins from 12 bonds $B$ neighboring the cube's edges:
\begin{equation}
  \begin{split}
    &H=-\sum_C\left[H_C\otimes\prod_{P\in \text{plaq}(\partial C)}H_P\otimes\prod_{B\in \text{bond}(\partial C)}H_B\right]\\
  &H_C=\ket{1}\bra{1}_C +\ket{1}\bra{0}_C+\ket{0}\bra{1}_C+\ket{0}\bra{0}_C,\\
  &H_P=\ket{1}\bra{1}_P+\ket{0}\bra{0}_P,\ \  H_B=\ket{1}\bra{1}_B+\ket{0}\bra{0}_B,
\end{split}
\label{eqn:3dczx_ham}
\end{equation}
where by $\ket{0}_C$ ($\ket{1}_C$) we denote a state which has all eight spins belonging to cube $C$ in the spin-down (spin-up) state, i.e., $\ket{0}_C\equiv \ket{00000000}_C$ ($\ket{1}_C\equiv \ket{11111111}_C$) and correspondingly for states of four spins on a plaquette $P,$ and states of two spins belonging to a bond $B$.
Each term in the Hamiltonian (\ref{eqn:3dczx_ham}) commutes with every other term, and the ground state is given by a tensor product over the states at each cube\cite{You18}:
\begin{equation}
  \ket{GS}=\bigotimes_C\frac{1}{\sqrt{2}}\left(\ket{1}_C+\ket{0}_C\right).
  \label{eqn:3dczx_GS}
\end{equation}

This model is a second order 3D topological phase, and has hinge modes protected by a combination of $C_4$ rotation symmetry and an on-site CZX symmetry group $\mathbb{Z}_2$. We expect therefore, that the ground state of the entanglement Hamiltonian will be unique and have the structure of a conventional (first-order) SPT state on the 2D entanglement cut. The reduced density matrix for a single cube that is split in half by the entanglement cut running parallel to one of its faces is given by:
\begin{equation}
  \rho=\frac{1}{2}(\ket{0000}\bra{0000}+\ket{1111}\bra{1111}),
  \label{eqn:rdm_cube}
\end{equation}
which yields an entanglement spectrum that is degenerate.
This goes against our naive expectations, but as one might also expect from the preceding discussions, this degeneracy exists because of subsystem symmetries. These symmetries are fine-tuned (not required to protect the HOSPT state), and can be broken by introducing perturbations preserving the required $C_4\times \mathbb{Z}_2$ symmetry.
Explicitly, we can add a set of plaquette terms on the plane $S$ neighboring (and inside) the entanglement cut:
\begin{equation}
  \begin{split}
  H_S=h\sum_{P\in S}\left(\ket{1}\bra{1}_P
  +\ket{0}\bra{1}_P+\ket{1}\bra{0}_P
  +\ket{0}\bra{0}_P\right).
\end{split}
  \label{eqn:surf_terms}
\end{equation}
For small $h$ these terms modify the entanglement Hamiltonian perturbatively, in a manner that gaps the entanglement Hamiltonian.
The resulting entanglement ground state is unique and has the following form at each plaquette on the entanglement surface:
\begin{equation}
  \ket{\Psi}^E=\bigotimes_{P\in S}\left(\ket{1}_P+\ket{0}_P\right).
  \label{eqn:CZX_ent_GS}
\end{equation}
Importantly, this is the 2D SPT ground state of the conventional CZX model\cite{Chen2011}, and represents the 2D SPT phase one will obtain at a physical edge after breaking subsystem symmetries. Now, we can take the next step and perform a second entanglement cut, this time on the entanglement ground state. The cubes in the bulk that are cut, i.e., those away from the first entanglement surface, contribute to a gapped second order entanglement spectrum. However, the second entanglement cut splits some of the \emph{plaquettes} of the 2D CZX model, and hence yields a second order NEH $H^{E,2}$ that is doubly degenerate and does not posses any residual symmetries other than those required to protect the HOSPT phase of the 3D CZX model. For this model we were able to directly identify that the entanglement ground state is a first-order SPT state. For more generic 3D Hamiltonians that exhibit the same 3D HOSPT phase one may have to consider more sophisticated properties of the entanglement spectrum to identify the HOSPT since (i) the entanglement ground state may not be the full story\cite{chandran2014}, and (ii) degeneracies of the nested entanglement cut of an effectively 2D entanglement surface may be spurious\cite{williamson2019} and require further analysis to identify the possible SPT state living on the initial entanglement cut.

\section{Conclusion}\label{sec:conclusion}
In this article we proposed using entanglement as a way to access the higher-order boundary physics of HOSPTs from the bulk wave functions. Our characterization method can be applied to many-body systems, and employs a series of \emph{nested} entanglement Hamiltonians that can be used to recover the topological properties of some HOSPTs. 
We have shown that it is possible to characterize a class of \emph{multipolar} HOSPT phases in both, interacting and non-interacting models using the entanglement structure of their ground state.
We considered a series of models falling into this category and showed that the sequence of nested entanglement Hamiltonians of order $1...(n-1)$ are gapped, while the $n$-th nested entanglement Hamiltonian is gapless. In our considerations we always found that an $n$-th order multipolar HOSPT phase has a gapped entanglement Hamiltonian that hosts an $(n-1)$-th HOSPT phase at the entanglement edge. We also considered a series of HOSPT phases having additional subsystem symmetry. In order to apply our method we had to break subsystem symmetry (effectively gapping out the boundaries). After breaking the subsystem symmetries, and leaving the global symmetry that protects the HOSPT untouched, we demonstrated that the corresponding nested entanglement Hamiltonian method could be applied and that these models fall into the category of multipolar-like HOSPT phases.

Our proposed algorithm that utilizes higher order entanglement spectra with a hierarchy of spatial cuts provides us with a powerful characterization method to probe HOSPT phases without referring to details of boundary terminations, but simply relying on the bulk ground state wavefunction. This could help pave the way for understanding the crucial many-body features of some classes of HOSPT phases. It also shows the necessity for further refinement of many-body HOSPT indicators as we expect that there will be difficulties when trying to extract universal information about HOSPTs from more generic models, and furthermore, the specific method we apply here may not work for all types of HOSPTs, e.g., those that fall out of the multipole paradigm. This will be an exciting direction for future work.

{\bf{Note:}} During the preparation of this manuscript we became aware of a related overlapping work "Higher-Order Entanglement and Many-Body Invariants for Higher-Order Topological Phases" by Yizhi You, Julian Bibo, and Frank Pollmann. These works were completed independently.

\textbf{Acknowledgements.} We thank Yizhi You for insightful comments on our manuscript and for informing us of her unpublished work. OD and TLH thank the US National Science Foundation for support under the award DMR 1351895-CAR, and the NSF MRSEC program under NSF Award Number DMR-1720633 (SuperSEED).

\bibliography{lib.bib}

\appendix

\section{Second order SPT phase without a bulk quadrupole moment.} \label{app:fqti}
To provide a contrasting free-fermion example to the QTI, let us consider an interesting variation of the QTI model which hosts modes at the physical corner, but was shown to develop a ground state that is qualitatively distinct from the ground state of the QTI model.
This model is constructed from the QTI model by removing the $\pi$-flux going through the plaquettes and slightly offsetting the strength of hopping amplitudes in different directions. 
On a rectangular-shaped lattice, this model is gapped both in its bulk and at its one-dimensional edges, and hosts gapless modes in the corners of the lattice.
The distinction between the ground states of these Hamiltonians can be directly seen from the nested Wilson loop construction\cite{quadrupole,quadrupole2}, here however, we will briefly show that this difference is also picked up by our nested entanglement Hamiltonian construction. Once again, let us work in the zero-correlation length limit with the inter-cell couplings offset: $\lambda_y>\lambda_x$. The analysis here will be exactly analogous to the one performed in the previous discussions on the free-fermion QTI. 
After making the first cut that runs along $\hat{y}$, as shown in Fig. \ref{fig:quadrupole_spectra}, the total entanglement Hamiltonian takes a familiar form:
\begin{equation}
    H^{E,1}_{AB}=\bigotimes_{i\in \mathcal{C}_{AB}}H_{\square,i}\bigotimes_{j\in \mathcal{C}_{AB\cap C}}H^{E,1}_{AB,j},
\end{equation}
where $H_{\square,i}=\lambda_x c^\dagger_{1,i} c_{2,i}+\lambda_y c^\dagger_{2,i} c_{3,i}+ \lambda_x c^\dagger_{3,i} c_{4,i}+\lambda_y c^\dagger_{4,i} c_{1,i}$ and the entanglement Hamiltonian for a cluster $j$ lying directly across the cut is $H^{E,1}_{AB,j}=\log(4)\mathbb{I}_{4\times 4}$ in the basis $\{\ket{00}_j,\ket{01}_j,\ket{10}_j,\ket{11}_j\}$. The ground state of the first entanglement Hamiltonian is thus massively degenerate rendering it impossible to unambiguously define the second-order NEH.

\section{Entanglement edge of the 2D XY model}
\label{app:ent_edge_2d_heis}

In this appendix we provide a detailed derivation of the first order correction to the entanglement Hamiltonian for the bosonic XY model with intra-cell couplings turned on, albeit kept perturbatively small. We start with the set of terms in the direct vicinity of a straight horizontal cut as shown in Fig. \ref{fig:line_x}:
\begin{equation}
  H^{cut}=\lambda_x H^X+\lambda_y H^Y+tH_A+tH_B,
  \label{eqn:heis2d1}
\end{equation}
where $H^{X}$ $H^{Y}$ are the collections of all inter-cell coupling terms oriented in the $\hat{x}$ and $\hat{y}$ directions respectively, while $H_A$ and $H_B$ are the intra-cell coupling terms which are entirely located either in the $A$ or $B$ subregions of the lattice.
For $t=0$ the ground state can be written as a tensor product of three states: one that is entirely contained in $A$, one that lies directly on the cut and one that is contained within the region $B$:
\begin{equation}
    \ket{GS}=\ket{GS}_A\otimes\ket{GS}_{AB}\otimes\ket{GS}_B.
\end{equation} 
In the zero correlation length limit, the ground state of this system remains in the non-trivial HOSPT phase even if we tune the values of the inter-cell couplings $\lambda_x$ and $\lambda_y$ to be very different, as such transformations do not break the $\mathbb{Z}_2\times\mathbb{Z}_2$ symmetry that protects the HOSPT phase, and they do not close the bulk gap. This can also be double-checked by computing the spectrum of the 2nd NEH for a single spin plaquette with, for example, $\lambda_y\gg \lambda_x$ and verifying that it is indeed doubly degenerate.

The entanglement cut running along the $\hat{x}$-direction slices through a set of $\lambda_y$ intra-cell couplings, as shown in Fig. \ref{fig:line_x}. Let us turn off every set of couplings in the system except the $H^{Y}$. We then compute the entanglement Hamiltonian for the system containing only $H^{Y}$ and then treat the rest of the couplings
\begin{equation}
    V=\lambda_x H^{X}+t(H_A+H_B)
\end{equation}
as a perturbation with the small parameter $\varepsilon$, so that the Hamiltonian under consideration is: $\tilde{H}^{cut}=\lambda_y H^Y+\varepsilon V$. 

The ground state $\ket{GS^{(0)}}$ of $H^{Y}$ is simply a collection of disjoint spin singlet states on each dimer. Let us consider the first-order correction to the ground state:
\begin{equation}
  \ket{\widetilde{GS}}=\ket{GS^{(0)}}-\varepsilon\sum_{k\neq 0}\ket{\psi_k}\frac{\bra{\psi_k}V\ket{GS^{(0)}}}{E_k-E_0}.
  \label{eqn:1st_order_pert}
\end{equation}
The energy gap between the ground state and every excited state connected to $\ket{GS^{(0)}}$ by any single term in $V$ is the same:  $E_k-E_0\equiv\Delta,$ and since $\bra{GS^{(0)}}V\ket{GS^{(0)}}=0$ we simply have:
\begin{equation}
  \ket{\widetilde{GS}}=\ket{GS^{(0)}}-\frac{\varepsilon}{\Delta}V\ket{GS^{(0)}}.
  \label{eqn:1st_order_pert_cont}
\end{equation}
We can use this first-order corrected state to compute the corresponding density matrix:
\begin{equation}
  \tilde{\rho}=\rho^{(0)}-\frac{\varepsilon}{\Delta}(V\rho^{(0)}+\rho^{(0)}V).
  \label{eqn:1st_order_rho}
\end{equation}
To compute the reduced density matrix $\tilde{\rho}_A$, we need to trace out the degrees of freedom on subregion $B$. 
Note that every single coupling term in $V$ has the form $\sigma^a_i\sigma^a_j$ with two spins $i$ and $j$ belonging to different singlets. As those terms do not have a projection back to the ground state, we find that out of all the coupling terms that are contained in region $B,$ the non-trivial contribution to $\tilde{\rho}_A$ will be given by only those terms that act on a pair of dimers directly at the entanglement cut. As one can check, terms in region $B$ that are aligned directly at the cut have exactly the same matrix elements as their counterparts that are mirrored on the `$A$' side of the cut. This allows us to rewrite the set of terms in $V$ that have a non-trivial contribution to $\tilde{\rho}_A$ in a way that does not have any explicit action on region $B$:
\begin{equation}
    \begin{split}
        &H^{X}=H^{X}_A+H^{X}_{\partial A};\quad H_{B}=H_{\partial A},
    \end{split}
\end{equation}
where $H^X_A$ is the set of all inter-cell couplings oriented along $\hat{x}$ that are entirely contained in $A,$ $H^{X}_{\partial A}$ is a subset of inter-cell couplings along $\hat{x}$ that lie at the entanglement cut, and each term in $H^{X}_{\partial A}$ acts simultaneously at two singlets crossing the cut; similarly for $H_{\partial A}$. 
Note that we have already thrown away terms that do not contribute to $\tilde{\rho}_A$. The relevant part of $V$ is then:
\begin{equation}
  V^*=\lambda_xH^{X}_A+\lambda_x H^{X}_{\partial A}+t\tilde{H}_{A}+tH_{\partial A}.
  \label{eqn:V_new}
\end{equation}
The resulting $V^*$ acts trivially on the subregion $B$. Therefore, after taking the partial trace over $B$, in the first order of the perturbation theory we have:
\begin{equation}
  \tilde{\rho}_A=\rho_A^{(0)}-\frac{\varepsilon}{\Delta}\{V^*,\rho_A^{(0)}\}.
  \label{eqn:rhoA_1st_order}
\end{equation}

To obtain the corresponding entanglement Hamiltonian, we need to take a $\log$ of this expression.
Using the analog of the Baker-Campbell-Hausdorff formula for the anticommutator, we find that the entanglement Hamiltonian along with the first order correction is simply given by:
\begin{equation}
  \tilde{H}^E_A=H^{E(0)}_A-2\frac{\varepsilon}{\Delta}V^*.
  \label{eqn:hamE_1st_order}
\end{equation} 
$H^{E(0)}_A$ is easy to compute directly: it is simply a set of dimers in the bulk of the subregion $A,$ just as in the original Hamiltonian $H^{Y},$ along with the set of free spin-1/2 degrees of freedom right beside the cut. Clearly then, the sum $H^{E(0)}_A-2\frac{\varepsilon}{\Delta}V^*$ represents a Hamiltonian of a bosonic XY HOSPT model with a physical edge right where we drew the entanglement cut.
Therefore, the first-order corrected entanglement Hamiltonian describes the original 2D XY model with a physical edge which is a 1D SPT phase\cite{Dubinkin18}.

\section{Topological Plaquette Paramagnet model}
\label{app:2d_tpp}
\textbf{Ground state of a 2D TPP model}
Consider the TPP model on a periodic lattice with a horizontal cut as shown in Fig. \ref{fig:2d_tpp}. Let us now divide our cylinder in two halves $A$ and $B$ by making a cut which runs diagonally with respect to primitive vectors of the lattice as shown in Fig. \ref{fig:2d_tpp}. Let us consider first a physical cut: we simply drop any stabilizers that are not fully supported on $A$ out of Eq. (\ref{eqn:Ham_z2z2_2}). This yields a system with a boundary, which, as was explored in Ref. \onlinecite{You18}, has a highly degenerate ground state. We can see this by noticing that we can associate a pair of anticommuting operators with each site at the edge, that all commute with the Hamiltonian:
\begin{equation}
  P^1_{i}=\tau^z_{i-1}\sigma^x_i\tau^z_{i+1},\ P^2_{edge}=\sigma^z_i.
  \label{eqn:edge_ops}
\end{equation}
Gapless modes associated with these operators are protected by a set of subsystem symmetries generated by $\prod_{i\in diag}\sigma^x_i$ and $\prod_{i\in diag}\tau^x_i$. We can break these symmetries by adding a potential term to the Hamiltonian:
\begin{equation}
  \varepsilon V=-\varepsilon\sum_{i\in edge}P^1_i=-\varepsilon\sum_{i\in edge}\tau^z_{i-1}\sigma^x_i\tau^z_{i+1}.
  \label{eqn:potential_term}
\end{equation}
On one hand, this term lifts the ground state degeneracy leaving a unique ground state in the system with open boundaries. On the other hand, it commutes with the symmetries that protect the HOSPT phase in our system.

Let us now, instead of a physical cut, make an entanglement cut and study its properties. First of all we need to write down the ground state of our system. Since we are working with a stabilizer Hamiltonian (having stabilizers denoted by $\mathcal{O}_i)$, our ground state must satisfy:
\begin{equation}
  \mathcal{O}_i\ket{\Psi^{(0)}}=\ket{\Psi^{(0)}},\ \forall \ \mathcal{O}_i\in H.
  \label{eqn:stab_GS}
\end{equation}
In our model we have two different types of stabilizers: ones that are centered around sites belonging to sublattice $a$, i.e., $\mathcal{O}^a_i=\tau^x_i\prod_{j\in P_i}\sigma^z_j$ where $i\in a$, and the others that are centered around sites belonging to $b$: $\mathcal{O}^b_i=\sigma^x_i\prod_{j\in P_i}\tau^z_j$ where $i\in b$. Hence, the Hamiltonian is just:
\begin{equation}
  H=-\sum_{i\in a}\mathcal{O}^a_i-\sum_{i\in b}\mathcal{O}^b_i.  \label{eqn:Ham_z2z2_2_stab}
\end{equation}

Similar to Kitaev's toric code, we can express the ground state of the system by starting with the ground state for one subset of stabilizers: the mutual ground state for stabilizers $\mathcal{O}^b$ is easy to find: we need all of the `red' spins to be `up' in the $\hat{x}$ basis, while all of the `blue' spins to be `up' in the $\hat{z}$ basis, so we write:
\begin{equation}
\ket{0}=\bigotimes_{i\in a}
\frac{1}{\sqrt{2}}\left(\begin{matrix} 1 \\ 1 \end{matrix}\right)_i
\bigotimes_{j\in b}
\left(\begin{matrix} 1 \\ 0 \end{matrix}\right)_j.
  \label{eqn:0_state}
\end{equation}
Now we form an abelian group $G$ with generators being $\mathcal{O}^a$ stabilizers and the ground state can be written as:
\begin{equation}
  \ket{\Psi^{(0)}}=\frac{1}{|G|}\sum_{g\in G}g\ket{0}.
  \label{eqn:gs_group}
\end{equation}
When one of the stabilizers $\mathcal{O}^a_i$ acts on a state $\ket{0}$, it flips the corresponding $\tau^z_i$ as well as its surrounding $\sigma^x$'s. 
Heuristically, we can think of the ground state as a superposition of all possible configurations of $\{\tau^z_i\}$ with corners of domain walls between different $\tau^z$ decorated with $\sigma^x_i=-1$.

\textbf{Entanglement edge of the 2D TPP model with broken subsystem symmetries}
Let us compute the first order correction to the entanglement Hamiltonian in the presence of the perturbation (\ref{eqn:potential_term}):
\begin{equation}
  \varepsilon V=-\varepsilon\sum_{i\in b\in \partial A}\tau^z_{i-1}\sigma^x_i\tau^z_{i+1}.
  \label{eqn:ssbreaking_term2}
\end{equation}
First, note that for every term in $V$ a state $\ket{\Psi_i}=\tau^z_{i-1}\sigma^x_i\tau^z_{i+1}\ket{\Psi^{(0)}}$ is an eigenstate of the Hamiltonian $H_{TPP}$ given by Eq. \ref{eqn:Ham_z2z2_2}. The energy of $\ket{\Psi_i}$ relative to the ground state $\ket{\Psi^{(0)}}$ (\ref{eqn:gs_group}) is $\Delta=E_i-E_0=8$, as there are exactly four stabilizers in $H_{TPP}$ that anticommute with the $\tau^z_{i-1}\sigma^x_i\tau^z_{i+1}$ operator. With this in mind we can, similarly to the perturbed 2D XY model case considered above, write down the first-order correction to the ground density matrix:
\begin{equation}
  \tilde{\rho}=\rho^{(0)}-\frac{\varepsilon}{\Delta}(V\rho^{(0)}+\rho^{(0)}V).
  \label{eqn:1st_order_rho_TPP}
\end{equation}
Since $V$ acts non-trivially only on the subregion $A$, it is left untouched when we take a partial trace over the subregion $B$. The perturbed reduced density matrix then reads:
\begin{equation}
  \tilde{\rho}_A=\rho_A^{(0)}-\frac{\varepsilon}{\Delta}(V\rho_A^{(0)}+\rho_A^{(0)}V).
  \label{eqn:rhoA_1st_order_TPP}
\end{equation}
As was the case with the 2D XY model, we use the anticommutator version of the Baker-Campbell-Hausdorff formula to take the logarithm of the expression for $\tilde{\rho}_A$ and obtain the perturbed entanglement Hamiltonian: 
\begin{equation}
  \tilde{H}^E_A=H^{E}_A-2\frac{\varepsilon}{\Delta}V.
  \label{eqn:hamE_1st_order_TPP}
\end{equation} 
With the perturbative correction taken into account we can see that the entanglement edge Hamiltonian is the Hamiltonian of a well-known $\mathbb{Z}_2\times\mathbb{Z}_2$ SPT chain.

\textbf{3D TPP model.}
This model is a natural extension of the 2D TPP model discussed in the previous paragraphs to 3D. It is defined on a body centered cubic lattice with the following Hamiltonian:
\begin{equation}
  H=-\sum_{i\in a}\tau^x_i\prod_{j\in C_i}\sigma^z_j-\sum_{i\in b}\sigma^x_i\prod_{j\in C_i}\tau^z_j,
  \label{eqn:Ham_z2z2_3D}
\end{equation}
where $a$ are the sites of the cubic lattice, $b$ are the sites of the dual lattice, and $C_i$ is the cube which has the site $i$ as its center. 
Let us write down the ground state in the $\tau^z$ and $\sigma^x$ basis. 
Starting with the state $\ket{0}$ that has all $\tau^z=+1$ and all $\sigma^x=+1$ we introduce a group $G$ generated by the set of stabilizers $\mathcal{O}_i=\tau^x_i\prod_{j\in C_i}\sigma^z_j$, where index $i$ belongs to the $a$ sublattice. The ground state is then given by:
\begin{equation}
    \ket{\Psi^{(0)}}=\frac{1}{|G|}\sum_{g\in G}g\ket{0}.
\end{equation}
Let us make an entanglement cut perpendicular to the $\hat{z}$ axis by a plane that lies slightly above a plane which has sublattice $a$ sites only and splits the lattice into $A$ (bottom) and $B$ (top) regions.
Tracing out the top region to form a reduced  density matrix, we find a contribution to the entanglement Hamiltonian that is localized right on the cut:
\begin{equation}
    H^{E,1}_{\partial A}=-\sum_{i\in a\in\partial A}\tau^x_i\prod_{j\in P_i}\sigma^z_j.
\end{equation}
This cut-localized Hamiltonian is a 2D commuting-projector model that has a large ground state degeneracy. We can see the degeneracy by noticing that to every site $i\in b\in \partial A$ we can associate a pair of anti-commuting operators that perfectly commute with the cut-localized entanglement Hamiltonian:
\begin{equation}
    Q_{i,1}=\sigma^x_i\prod_{j\in P_i}\tau^z_j,\,\,\,\ Q_{i,2}=\sigma_i^z. 
\end{equation}
The gapless modes encoded by this algebra are protected by subsystem symmetry generated by the operators $\prod_{i\in a\in \text{column}}\tau^x$.
Therefore, to proceed with our calculation to arrive at a second order entanglement spectrum, we need to break these subsystem symmetries and gap out the cut-localized entanglement  Hamiltonian. We can do so by using the same approach as in the 2D TPP model, e.g., by adding the set of terms 
\begin{equation}
    \Delta H=\sum_{i\in b\in\partial A}\sigma^x_i\prod_{j\in P_i}\tau^z_j.
\end{equation}
Indeed, we see that the total edge Hamiltonian $H^{E,1}+\Delta H$ is nothing but a 2D TPP model, which, was already shown to be a quadrupolar HOSPT phase augmented by subsystem symmetries.
Thus, we see that the third order NEH $H^{E,3}$ of the 3D TPP model will have a doubly-degenerate eigenspectrum after we break all the necessary subsystem symmetries.

\section{XY cluster}
Let us consider a cluster of four spins with the following Hamiltonian:
\begin{equation}
  H=\sum_{a=x,y}\sigma^a_1\sigma^a_2+\sigma^a_2\sigma^a_3+\sigma^a_3\sigma^a_4+\sigma^a_4\sigma^a_1.
  \label{Heis_cluster}
\end{equation}
The ground state of this Hamiltonian is:
\begin{equation}
\begin{split}
  \ket{\psi^{(0)}}=&\frac{1}{2\sqrt{2}}\left(\ket{\uparrow\uparrow\downarrow\downarrow}+\ket{\downarrow\uparrow\uparrow\downarrow}+\ket{\downarrow\downarrow\uparrow\uparrow}+\right.\\
  &\left.\ket{\uparrow\downarrow\downarrow\uparrow}\right)-\frac{1}{2}\left(\ket{\downarrow\uparrow\downarrow\uparrow}+\ket{\uparrow\downarrow\uparrow\downarrow}\right).
\end{split}
\end{equation}
Now, let us compute the reduced density matrix for the entanglement cut that eliminates spins $3$ and $4$ and splits the initial plaquette into a pair of dimers. The reduced density matrix is then:
\begin{equation}
  \rho_{\{1,2\}}=\ket{v_1}\bra{v_1}+\ket{v_2}\bra{v_2}+\ket{v_3}\bra{v_3}+\ket{v_4}\bra{v_4},
  \label{rho12}
\end{equation}
where:
\begin{equation}
  \begin{split}
    &\ket{v_1}=\frac{1}{2\sqrt{2}}\ket{\uparrow\uparrow},\ \ket{v_2}=\frac{1}{2\sqrt{2}}\ket{\downarrow\uparrow}-\frac12\ket{\uparrow\downarrow},\\
    &\ket{v_3}=\frac{1}{2\sqrt{2}}\ket{\uparrow\downarrow}-\frac12\ket{\downarrow\uparrow},\ \ket{v_4}=\frac{1}{2\sqrt{2}}\ket{\downarrow\downarrow}.
  \label{v1v2v3v4}
\end{split}
\end{equation}
In the basis $\{\ket{\downarrow\downarrow},\ket{\downarrow\uparrow},\ket{\uparrow\downarrow},\ket{\uparrow\uparrow}\}$ the reduced density matrix takes the following form:
\begin{equation}
  \rho_{\{1,2\}}=\frac18\left(\begin{matrix}
      1 & 0 & 0 & 0\\
      0 & 3 & -2\sqrt{2} & 0\\
      0 & -2\sqrt{2} & 3 & 0\\
      0 & 0 & 0 & 1
  \end{matrix}\right),
  \label{rho12num}
\end{equation}
and the entanglement spectrum is given by: $\{\frac18(3-2\sqrt{2}),\frac18,\frac18,\frac18(3+2\sqrt{2})\}$ and therefore the entanglement Hamiltonian has a unique ground state given by: 
\begin{equation}
    \ket{\psi^{(1)}}=\frac{1}{\sqrt{2}}\left(\ket{\uparrow\downarrow}-\ket{\downarrow\uparrow}\right).
\end{equation}
This is the easily recognizable ground state of a single physical XY dimer, and so it is evident that the second nested entanglement Hamiltonian will have a doubly degenerate spectrum as the reduced density matrix is simply: 
\begin{equation}
    \rho^{(2)}_{\{1\}}=\frac12\left(\begin{matrix}
	1 & 0\\
	0 & 1
  \end{matrix}\right),
\end{equation}
meaning that the eigenspectrum of $H^{E,2}=-\log(\rho^{(2)}_{\{1\}})$ is simply $\{\log(2),\log(2)\}$.
$$
\quad
$$

\end{document}